\documentclass[10pt,showpacs,floatfix,letterpaper,amsmath,amsfonts,amssymb,aps,pra,reprint,superscriptaddress]{revtex4-1}
\usepackage[latin1]{inputenc}
\usepackage{bm}
\usepackage{graphicx}
\usepackage{tensor}
\usepackage{epstopdf}
\usepackage{enumerate}
\newcommand{\ket}[1]{\lvert #1\rangle}                        % Ket
\newcommand{\bra}[1]{\langle #1 \rvert}                       % Bra
\newcommand{\braket}[2]{\langle #1 \vert #2 \rangle}          % Bra-Ket
\newcommand{\pr}[1]{\ket{#1}\bra{#1}}                         % Projector
\newcommand{\Tr}[1]{\text{Tr}(#1)}                            % Trace
\begin{document}
\title{Implementing generalized measurements with superconducting qubits}
\author{Justin Dressel}
\affiliation{Department of Electrical Engineering, University of California, Riverside, CA 92521, USA.}
\author{Todd A. Brun}
\affiliation{Communication Sciences Institute, University of Southern California, Los Angeles, CA 90089, USA}
\author{Alexander N. Korotkov}
\affiliation{Department of Electrical Engineering, University of California, Riverside, CA 92521, USA.}

\date{\today}

\begin{abstract}
We describe a method to perform any generalized purity-preserving
measurement of a qubit with techniques tailored to superconducting
systems. First, we consider two methods for realizing a two-outcome
partial projection: using a thresholded continuous measurement in
the circuit QED setup, or using an indirect ancilla qubit
measurement. Second, we decompose an arbitrary purity-preserving
two-outcome measurement into single qubit unitary rotations and a
partial projection. Third, we systematically reduce any
multiple-outcome measurement to a sequence of such two-outcome
measurements and unitary operations.  Finally, we consider how to
define suitable fidelity measures for multiple-outcome generalized
measurements.
\end{abstract}

%\pacs{42.50.Xa,03.65.Ta,42.50.Dv}
%\pacs{03.65.Ta,03.67.-a,02.50.Cw,03.65.Fd}

\maketitle

\section{Introduction}
The ability to isolate and control coherent quantum systems has
dramatically improved in recent decades.  As a consequence,
techniques that were previously restricted to thought experiments
have recently been promoted to practical laboratory methods.
Generalized quantum measurements fall into this category.

The concept of measurement in quantum mechanics has been largely
dominated historically by projective measurements, in which an
experimenter learns precise information about a quantum system under
study \cite{Dirac1930}.  Such projective measurements are commonly
called ``von Neumann measurements'' since von Neumann first
formalized the quasi-Boolean lattice of projection operators used to
describe the curious logic obeyed by these measurements
\cite{VonNeumann1932,Jauch1968}.  However, it is worth noting that
von Neumann simultaneously introduced the possibility of learning
imprecise information about a quantum system by measuring a
correlated ancilla that acts as an indirect detector for the system
\cite{VonNeumann1932}.  When such an indirect detector becomes
correlated with the system, an experimenter still obtains some
information about the system, but the precision of that information
depends on the degree of correlation between the ancilla and the
system.  These imprecise measurements are an example of generalized
measurements.

The mathematical description of generalized quantum measurements has
been considerably refined since these early observations
\cite{Davies1976,Holevo1982,Kraus1983,Aharonov1988,Braginski1992},
giving rise to the modern formalism of \emph{quantum operations}.
This formalism has proven itself invaluable for reasoning about
tasks in quantum information and quantum computing
\cite{Nielsen2000,Keyl2002,Wiseman2009,Korotkov-01,Dressel2012b}.
However, it is only in the past few decades that experimental
systems have become sufficiently controllable to make generalized
measurements a practical laboratory tool.

Thus far, optical systems have been the primary arena for
implementing generalized quantum measurements, with experiments
using them to examine nonorthogonal state discrimination
\cite{Huttner1996}, non-destructive photon measurements
\cite{Pryde2004,Guerlin2007,Gleyzes2007}, feedback control
\cite{Sayrin2011,Zhou2012b}, entanglement distillation
\cite{Kwiat2001}, weak value super-oscillation effects
\cite{Ritchie1991,Pryde2005,Iinuma2011}, Leggett-Garg inequality
violations \cite{Goggin2011,Dressel2011,Suzuki2012}, weak-value
amplification
\cite{Hosten2008,Dixon2009,Starling2010b,Turner2011,Zhou2012,Gorodetski2012,Zhou2013},
locally-averaged photon trajectories \cite{Kocsis2011}, direct
wave-function determination \cite{Lundeen2011}, error-disturbance
complementarity \cite{Rozema2012,Kaneda2013}, conditional
measurement reversal \cite{Kim2009,Kim2012}, Hardy's paradox
\cite{Lundeen2009,Yokota2009}, and much more.  In addition to these
specific examples of generalized measurements, there have been
several discussions about how to implement any desired measurement
on optical qubits \cite{James2001,Andersson2008,Ota2012}.
Nevertheless, current optical qubit architectures are not easily
scalable.  While measuring one or two photonic qubits in a general
way is possible using linear optics and parametric down-conversion,
it is not so easy to reliably entangle and manipulate larger numbers
of independent photonic qubits.  This scaling difficulty limits the
potential applications of generalized measurements.

In contrast, solid-state systems have demonstrated better
scalability in recent years, and have also implemented a variety of
generalized quantum measurements.  Experiments with superconducting
qubits have used generalized measurements to demonstrate partial
collapse \cite{Katz2006} and measurement reversal \cite{Katz2008},
violate Leggett-Garg inequalities
\cite{Palacios-Laloy2010,Groen2013}, stabilize Rabi oscillations
with quantum feedback control \cite{Vijay2012}, demonstrate quantum
back-action in an individual continuous measurement
\cite{Hatridge2013}, observe single quantum trajectories
\cite{Murch2013,Weber2014}, entangle qubits by measurement
\cite{Riste2013,Roch-14}, and reduce decoherence via uncollapsing
\cite{Zhong2014}. These systems show promise for realizing scalable
architectures that can manipulate many entangled qubits
simultaneously.  As such, we expect that many more applications of
generalized measurements will soon appear.  It is thus of particular
interest to specify exactly how to implement generalized
measurements with these systems in a systematic way \cite{Ota2012}.

In this paper, we discuss how to use modern superconducting systems
to perform any generalized qubit measurement. Our strategy will be
to reduce an arbitrary $k$-outcome measurement to a more manageable
sequence of two-outcome measurements.  Each such two-outcome
measurement can be further decomposed into single-qubit unitary
rotations and a standardized partial projection.  We discuss two
different ways to realize such a partial projection, specifying
explicitly the experimentally-controllable parameters. For
simplicity of discussion, we consider efficient detectors in what
follows, with the understanding that laboratory implementations will
have imperfect fidelity in practice. As such, we also examine the
issue of how to characterize the fidelity of a many-outcome generalized
quantum measurement.

The paper is organized as follows.  In
Section~\ref{sec:implementation}, we detail two methods of
implementing two-outcome partial projections using superconducting
qubits.  In Section~\ref{sec:reduction} we describe how to implement
any generalized measurements by decomposing them into sequences of
unitary operations and partial projections. In Section
\ref{sec:fidelity} the ways to characterize the fidelity of a
generalized measurement are discussed (with more details in the Appendix).
We conclude in Section~\ref{sec:conclusion}.

\section{Two-outcome Partial Projections}\label{sec:implementation}

To understand how to implement an arbitrary generalized measurement,
we first consider how to implement the simple case of a two-outcome
partial projection.  We will then be able to construct the general
case as an appropriate sequence of these partial projections and
additional unitary rotations.

Recall that in a projective two-outcome qubit measurement, the state
collapses into an eigenstate of the measurement $\ket{\psi} \to
\ket{0},\ket{1}$.  By convention, these eigenstates correspond to
the Pauli $Z$ operator $\sigma_z = \pr{0} - \pr{1}$.  Formally,
the collapse can be understood as the application of a particular
projection operator, $\pr{0}$ or $\pr{1}$, followed by state
renormalization.  The measurement result of $0$ or $1$ determines
which projection operator is applied, and thus fully determines the
qubit state after the measurement. In the special case when the
initial state is either $\ket{0}$ or $\ket{1}$, the measurement
result is deterministic, perfectly correlated with the qubit state,
and measurement does not change the qubit state.

For a generalized two-outcome measurement, the measurement result
can be ``noisy''.  That is, the measurement result of $0$ or $1$
need not perfectly correlate with the pre-measurement qubit state.
Instead, the results will correspond to the qubit state only
probabilistically. (Note that we assume the simplest case here where
the measurement basis of $|0\rangle$ and $|1\rangle$ is unchanged).  If
the qubit is in state $\ket{0}$, then there is a probability $p$
that the result  will correctly report $0$. Similarly, if the qubit
is in state $\ket{1}$, then there is a (generally different) probability $q$
that the result will correctly report $1$.  Since the obtained
information is imperfect, the state will only partially collapse
toward the eigenstates of the measurement when a result is obtained.

The simplest form of such a partial collapse (which does not involve
unitary evolution and/or decoherence -- see
\cite{Korotkov2008} for more details) is described formally by
partial projection operators
\begin{align}\label{eq:standard}
  D_0 &= \sqrt{p}\, \pr{0} + \sqrt{1-q}\, \pr{1}, \\
  D_1 &= \sqrt{1-p}\, \pr{0} + \sqrt{q}\, \pr{1}, \nonumber
\end{align}
that depend on the two probability  parameters (measurement
fidelities) $p,q\in[0,1]$. When a result $0$ (or $1$) is obtained,
the qubit state is updated to $D_0\ket{\psi}$ (or
$D_1\ket{\psi}$) and then renormalized. The probabilities of
obtaining results 0 and 1 are $||D_0\ket{\psi}||^2$ and
$||D_1\ket{\psi}||^2$, respectively. It is natural to assume
$p+q\geq 1$ so that the stated correspondence between $D_0$
and $0$ (or $D_1$ and $1$) is sensible; however, we will not
need to enforce this assumption.

When $p=q=1$, the projective measurement is recovered as a special
case. The case $p=1$, $q\neq 1$ is often called a null-result
measurement for the outcome 0 (the outcome 1 collapses the state to
$|1\rangle$, while the outcome 0 produces only a partial collapse towards
$\ket{0}$). The case $p+q=1$
corresponds to no measurement, with $p$ and $q$ directly indicating the
probabilities of the results 0 and 1, independently of the qubit state.
Thus, $|p+q-1|$ characterizes the strength of the measurement (i.e.,
how well the measurement can discriminate between the states $|0\rangle$ and
$|1\rangle$); for example, a ``weak'' measurement (in the sense of Ref.\ \cite{Aharonov1988}) satisfies $|p+q-1|\ll 1$.
The difference $p-q$ characterizes the asymmetry between the
probabilities of results 0 and 1; in particular, the probability of
the result 0 averaged over any qubit state is $(1+p-q)/2$, while the
averaged probability of the result 1 is $(1+q-p)/2$.

We now consider two methods available for superconducting qubits to
implement such a partial projection for an arbitrary choice of $p$
and $q$.

\subsection{Thresholded Continuous Readout}

The standard readout of a superconducting qubit in the circuit QED
setup
\cite{Blais2004,Wallraff2004,Chow2011,Johnson2012,Riste2012,Barends2013,Hatridge2013,
Palacios-Laloy2010,Groen2013,Vijay2012,Hatridge2013,Murch2013,Weber2014,Riste2013,Roch-14} involves a quadrature
measurement of the leaked output from a pumped microwave resonator
that is dispersively coupled to the qubit.
  In this case the qubit state evolves stochastically in the process
of its continuous measurement \cite{Gambetta2008,Korotkov2011} (see
also \cite{Wiseman2009,Korotkov-01,Clerk2010}).

Let us assume that a quadrature at an angle $\alpha$ from the
information-carrying quadrature is amplified by a quantum-limited
phase-sensitive amplifier and then measured. The instantaneous
output signal (which includes noise) is denoted as $I(t)$. The
average of this noisy signal is correlated with the state of the
qubit, so that the dimensionless readout $r(t)$ that averages to the
$\sigma_z$ range of $[-1,1]$ is
\begin{align}
  r(t) &= 2\, \frac{I(t) - I_c}{\Delta I},
\end{align}
where $I_c = (I_0 + I_1)/2$, $\Delta I = I_0 - I_1$, and the values
$I_0$ and $I_1$ are the average signals obtained when the qubit is
fixed in the states $|0\rangle$ and $|1\rangle$, respectively. The readout
$r(t)$ corresponds to the $z$-component of the qubit state on the
Bloch sphere. Note that $I_c$ will depend on the
quadrature angle $\alpha$ in general.  The response
$\Delta I$ will also depend on $\alpha$ as $\Delta I=\Delta I_{\rm max}\cos
\alpha$, where $\Delta I_{\rm max}$ is the maximum response
at angle $\alpha =0$.

In the quantum non-demolition (QND) regime \cite{Braginski1992} with
no additional unitary evolution, the integrated readout
\begin{align}\label{eq:readout}
  R = \int_0^T\frac{dt}{\tau}\,r(t)
\end{align}
completely determines the partial projection of the qubit
\cite{Korotkov-01,Korotkov2011}. Here $\tau = 2 S / (\Delta I)^2$ is
a characteristic ``measurement time'' that controls the rate of
partial projection (the signal-to-noise ratio of 1 is reached after
time $\tau$), while $S$ is the (approximately constant) one-sided
spectral density of the noisy signal due to the amplifier noise,
which is assumed here to be quantum-limited.
 [The variance of $I(t)$ is related
to the spectral density as $\text{Var}(I) = \int_0^\infty S(\omega)
\, d\omega/2\pi$.]
%The measurement time $\tau$ can also be obtained directly as the two-sided spectral density of the dimensionless readout $r$
%\[
%  \text{Var}(r) = \lim_{T\to\infty}\int_0^T\!\!\frac{dt}{T}r^2(t) = \int_{-\infty}^\infty \frac{d\omega}{2\pi}\,\tau(\omega).
%\]
Notably, the integration duration $T$ in Eq.~\eqref{eq:readout} is
arbitrary, so one can wait for a desired integrated readout $R$ to
appear, and then terminate the measurement (i.e., stop pumping the
microwave resonator).

If the output of the microwave resonator is collected efficiently
(i.e., without quantum information loss in connectors, transmission
lines, and amplifying circuitry), then each integrated readout $R$
corresponds to a purity-preserving \cite{purity-pres} measurement
that is characterized by a partial-projection operator (see
\cite{Korotkov2011})
\begin{align}\label{eq:measop}
  M_R &\propto \exp\left[ \frac{R}{2\cos\alpha} e^{-i \alpha}
  \sigma_z\right]
    \\
 &= e^{R/2}e^{-i(R/2)\tan \alpha} |0\rangle \langle
0| + e^{-R/2}e^{i(R/2)\tan \alpha} |1\rangle \langle 1|. \nonumber
\end{align}
After state renormalization, the constant proportionality factor
will cancel. Non-zero $\alpha$ increases the typical measurement
timescale $\tau=\tau_{min}/\cos^2 \alpha$ and produces $z$-rotations
of the qubit state that depend on the integrated result $R$. For
simplicity in what follows, we will assume measurement of the optimal
quadrature, $\alpha=0$.

An experimenter can then follow a simple procedure to implement the
two-outcome partial projection in Eq.~\eqref{eq:standard}:
\begin{enumerate}
  \item Set a positive value $R_0$ and a negative value $R_1$ [given later
  in Eq.~\eqref{R0R1}]
as threshold values for the integrated readout $R$.
  \item Wait for one of the threshold values to appear and then
  terminate the measurement.
\end{enumerate}
According to Eq.~\eqref{eq:measop}, this procedure will produce one of the partial projections
\begin{align}\label{eq:contstand}
  D_{0,1} &= \sqrt{C_{0,1}}\left[ e^{R_{0,1}/2}\, \pr{0}
  + e^{-R_{0,1}/2}\, \pr{1} \right],
\end{align}
where the normalization constants $C_{0,1}$  can be obtained either
by the first-passage techniques similar to Refs.\
\cite{Kor-Jordan-06,Jordan-Kor-10} or simply by using the condition
$D_0^\dagger D_0 + D_1^\dagger D_1 = \openone$, which
follows from the fact that at least one of the two thresholds will
eventually be reached (in turn, this follows from the fact that at
infinite time our continuous measurement  would collapse the qubit
state to either $|0\rangle$ or $|1\rangle$).

 The thresholds $R_{0,1}$ will determine the probabilities $p$ and $q$
in Eq.~\eqref{eq:standard}. Squaring the operators in
Eq.~\eqref{eq:contstand} and comparing them to
Eq.~\eqref{eq:standard} produces the relations:
\begin{align*}
  D_0^\dagger D_0 &= C_0 \left[ e^{R_0}\, \pr{0} + e^{-R_0}\, \pr{1}\right], \\
          &= p\, \pr{0} + (1-q)\, \pr{1}, \\
          D_1^\dagger D_1 &= C_1\left[ e^{R_1}\, \pr{0} + e^{-R_1}\, \pr{1}\right], \\
          &= (1-p)\, \pr{0} + q\, \pr{1}.
\end{align*}
It follows by inspection that
\begin{align}
  C_0 &= \sqrt{p(1-q)}, & C_1 &= \sqrt{q(1-p)} \\
  \label{R0R1}
  R_0 &= \frac{1}{2}\ln\left(\frac{p}{1-q}\right), &
  R_1 &= - \frac{1}{2}\ln\left(\frac{q}{1-p}\right).
\end{align}

  Thus, Eq.~\eqref{R0R1} gives us the threshold values $R_0 > 0$ and
$R_1 < 0$ that need to be set to perform the partial projection in
Eq.~\eqref{eq:standard} with arbitrary $p$ and $q$. Note that measuring
a different quadrature angle $\alpha$ will require the same thresholds,
but will produce additional $z$-rotations that are absent in
Eq.~\eqref{eq:standard}.

This way of realizing the partial projection in the circuit QED
setup is a direct generalization of the ``uncollapsing''
measurements considered in Refs.~\cite{Kor-Jordan-06} and
\cite{Jordan-Kor-10}. This measurement technique can also be
viewed as an experimental realization of the
type of continuous measurement decomposition of a generalized
measurement described in
\cite{Oreshkov2005,Oreshkov2006,Varbanov2007}.  For two-outcome
measurements on a qubit, only minimal feedback is necessary---that
is, determining when the measurement process should terminate.  For
higher-dimensional systems similar continuous measurement
decompositions exist, but in general more sophisticated feedback is
needed in the measurement process.

From Eq.~\eqref{R0R1} it is easy to check that a projective  measurement ($p=q=1$) requires $R_0=-R_1=\infty$; such a complete measurement can
only be realized approximately. The null-result measurement ($p=1$, $q\neq 1$)
requires $R_1=-\infty$ and finite positive $R_0$, so that the
result 1 gives complete information, while the result 0 is
inconclusive. The case of no measurement ($p+q=1$) gives
$R_0=R_1=0$, which means that the measurement is immediately terminated.
A weak measurement ($|p+q-1|\ll 1$, with $p$ and $q$ not too close to 0 or 1) corresponds
to small values of the thresholds, $R_0\ll 1$ and $|R_1| \ll 1$, so that
the measurement procedure likely lasts for a short time. A symmetric
measurement ($p=q$) requires symmetric thresholds, $R_1=-R_0$.

There is a significant caveat to this partial projection
implementation: the operator $M_R$ in
Eq.~\eqref{eq:measop} strictly applies only for a purity-preserving
(i.e., efficient) measurement. Experimentally, a quadrature readout
typically has imperfect quantum efficiency, which causes
additional state decoherence during the readout \cite{Korotkov2011}.
For such an inefficient measurement, the duration $T$ of the
integrated readout will matter, since it will determine the
accumulated decoherence.  As such, the thresholding technique will
generally produce a fluctuating distribution of measurements with different
amounts of additional decoherence, and so will only approximate the
desired partial projection with some average fidelity.

\subsection{Ancilla Qubit Measurement}

As an alternative to thresholding a continuous dispersive readout,
one can also realize the partial projection in Eq.~\eqref{eq:standard} with
arbitrary $p$ and $q$ as a quantum circuit using an ancilla qubit
measurement.
  This method does not require a continuous measurement with perfect
quantum efficiency, but it does require high-fidelity two-qubit entangling
operations and single-qubit gates; it also requires high-fidelity
projective measurement of the ancilla qubit. This method can be
realized with various types of qubits (not necessarily in
circuit QED systems) and at present is easier to implement experimentally
for superconducting qubits (e.g., Refs. \cite{Groen2013,Zhong2014})
than the method discussed in the previous subsection.

  The procedure requires standard one-qubit and two-qubit gates with
adjustable parameters. In particular, the one-qubit gates we will
use are the $X$, $Y$, and $Z$-rotations around the Bloch sphere
\cite{Nielsen2000}
\begin{align*}
  R_x(\phi) &= e^{-i \phi \sigma_x / 2}
  = \begin{pmatrix}\cos\frac{\phi}{2} & -i\sin\frac{\phi}{2}
  \\  i\sin\frac{\phi}{2} & \cos\frac{\phi}{2}\end{pmatrix},
    \\
  R_y(\phi) &= e^{-i \phi \sigma_y / 2} =
  \begin{pmatrix}\cos\frac{\phi}{2} & -\sin\frac{\phi}{2}
  \\ \sin\frac{\phi}{2} & \cos\frac{\phi}{2}\end{pmatrix},
    \\
   R_z(\phi) &= e^{-i \phi \sigma_z / 2} =
   \begin{pmatrix}e^{-i\phi/2} & 0 \\ 0 & e^{i\phi/2}\end{pmatrix}.
\end{align*}
For most superconducting qubit implementations, $X$ and $Y$
rotations are realized with microwave pulses. The $Z$ rotation can
be realized either by changing the qubit frequency or as a
composition of $X$- and $Y$-rotations, e.g., $R_z(\phi) =
R_x(\pi/2)R_y(\phi)R_x(-\pi/2)$.
%Notice that the Hadamard gate is also a composition of two rotations and a global phase, $ H = e^{i\pi/2}R_x(\pi)R_y(\pi/2)$.

The partial projection procedure also requires a two-qubit
entangling gate.  The most convenient gate to use for conceptually
understanding a partial projection is a $Z$-controlled $Y$-rotation
(or $X$-rotation) of the form
\begin{align}\label{eq:zconty}
  R_{y|z}(\phi) &= e^{-i \phi (\sigma_z\otimes\sigma_y)/2} \nonumber \\
  &= \begin{pmatrix}\cos\frac{\phi}{2} & -\sin\frac{\phi}{2} & 0 & 0 \\ \sin\frac{\phi}{2} & \cos\frac{\phi}{2} & 0 & 0 \\
    0 & 0 & \cos\frac{\phi}{2} & \sin\frac{\phi}{2} \\ 0 & 0 & -\sin\frac{\phi}{2} & \cos\frac{\phi}{2}\end{pmatrix}.
\end{align}
This gate rotates the qubit in the $Z$-$X$ plane of the Bloch sphere
by an angle $\pm \phi$ depending on the state of the control qubit.
This gate may be produced directly if the qubit implementation
admits an effective $Z$-$Y$ (or $Z$-$X$) interaction Hamiltonian of
the form $\hbar\Omega(\sigma_z\otimes\sigma_y)$ (e.g.,
\cite{Chow2011,Chow2013}). Alternatively, as discussed later, it may
be realized by using a controlled-phase gate
\begin{align} \label{eq:C-phase}
  C\!Z(2\phi) &= \begin{pmatrix}1 & 0 & 0 & 0 \\ 0 & 1 & 0 & 0
  \\ 0 & 0 & 1 & 0 \\ 0 & 0 & 0 & e^{i 2\phi}\end{pmatrix},
\end{align}
that is properly dressed by $X_{\pi/2}$ rotations of the ancilla
(e.g., \cite{Groen2013}); the angle in Eq.~\eqref{eq:C-phase}
is $2\phi$ because the total phase difference is $2\phi$ in the gate
\eqref{eq:zconty}. Yet another way to realize the gate \eqref{eq:zconty}
is by using a fixed controlled-$Z$ gate $C\!Z(\pi)$ and one-qubit
rotations that depend on $\phi$ (as we shall see shortly).

\begin{figure}[t]
  \begin{center}
    \includegraphics[width=\columnwidth]{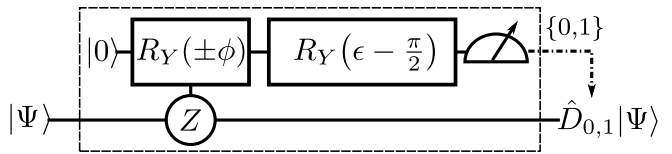}
  \end{center}
  \caption{Quantum circuit implementing the partial projections
  $D_{0,1}$ using an ancilla qubit.  The ancilla is initialized
  in the state $\ket{0}$ and then a $Z$-controlled $Y$-rotation
  of $\pm\phi$ is applied [see Eq.~\eqref{eq:zconty}], which creates
  an angular separation of $2\phi$ between the ancilla states
  coupled to the qubit states of $\ket{0}$ and $\ket{1}$.  Finally,
  the ancilla is $Y$-rotated by the angle $\epsilon-\pi/2$ and measured.
  For $\epsilon=0$ this corresponds to measurement in the $X$-basis; the
  offset $\epsilon$ allows for measurement asymmetry. The
  resulting partial projection of the qubit $D_{0,1}$ depends
  on the classical outcome $0$ or $1$ of the ancilla measurement.
  }
  \label{fig:circuit}
\end{figure}

Now let us discuss the protocol to implement the partial projection
of Eq.~\eqref{eq:standard}. Using the $Z$-controlled $Y$ rotation
\eqref{eq:zconty}, it can be done via the procedure illustrated in
Fig.~\ref{fig:circuit}:
\begin{enumerate}
  \item Initialize the ancilla qubit in the state $\ket{0}$.
  \item Perform a $Z$-controlled $Y$ rotation \eqref{eq:zconty}
  of the ancilla by an angle $\phi$ (to be determined later),
  which entangles the main qubit with the ancilla.
  \item Perform a $Y$ rotation of the ancilla
  $R_y(\epsilon-\pi/2)$, with the offset angle $\epsilon$ to be
  determined later.
  \item Measure the ancilla projectively in the computational basis
  $\{\ket{0},\ket{1}\}$.
\end{enumerate}

The idea behind this procedure is to first create two states of the
ancilla that correspond to the main qubit states $|0\rangle$ or
$|1\rangle$, and that are both in the $Z$-$X$ plane at angles $\pm
\phi$ from the $Z$ axis. These ancilla states are then measured along a
direction in the same $Z$-$X$ plane at an angle $\epsilon$
from the $X$ axis. The angle $\phi$ then determines the effective
distinguishability of the states $|0\rangle$ and $|1\rangle$ of the
main qubit: when $\phi=0$ the states are indistinguishable. The offset
angle $\epsilon$ introduces asymmetry between the averaged
probabilities of the measurement results, with $\epsilon =0$ indicating
perfect symmetry.

Quantitatively, the results $0$ or $1$ of the ancilla measurement
produce the following partial projections of the main qubit:
\begin{align} \label{eq:qubitpart}
  D_0 &= \bra{0}\,[\openone \otimes R_y(\epsilon-\pi/2)]\, R_{y|z}(\phi)\, \ket{0} \\
  &= \sqrt{\frac{1 + \sin (\phi + \epsilon)}{2}}\, \pr{0} +
  \sqrt{\frac{1 - \sin (\phi - \epsilon )}{2}}\, \pr{1}, \nonumber \\
  D_1 &= \bra{1}\,[\openone \otimes R_y(\epsilon-\pi/2)]\, R_{y|z}(\phi)
  \, \ket{0}  \\
  &= \sqrt{\frac{1 - \sin ( \phi + \epsilon )}{2}}\, \pr{0} +
  \sqrt{\frac{1 + \sin ( \phi - \epsilon )}{2}}\, \nonumber \pr{1}.
\end{align}

By comparing the form of Eq.~\eqref{eq:qubitpart} to the standard
form of Eq.~\eqref{eq:standard}, the parameters $p$ and $q$ are
\begin{align}
  p &= \frac{1}{2}\left[1 + \sin(\phi + \epsilon)\right], \\
  q &= \frac{1}{2}\left[1 + \sin(\phi - \epsilon)\right].
\end{align}
Therefore, any desired parameters $p$ and $q$ may be realized by
setting the angles of the implementation circuit in Fig.\
 \ref{fig:circuit} to
\begin{align}
  \phi = \frac{\arcsin(2p-1) + \arcsin(2q-1)}{2},
  \label{eq:phi}\\
  \epsilon = \frac{\arcsin(2p-1) - \arcsin(2q-1)}{2}.
  \label{eq:epsilon}
\end{align}

It is easy to check that projective measurement ($p=q=1$) requires
$\phi=\pi/2$ and $\epsilon =0$. The case of no measurement ($p+q=1$)
is realized when $\phi=0$, and a weak measurement ($|p+q-1|\ll 1$)
requires $|\phi|\ll 1$. The symmetric case ($p=q$) corresponds to
$\epsilon = 0$. A null-result measurement ($p=1$) is realized when
$\phi +\epsilon =\pi/2$, so that the qubit state $\ket{0}$ always
produces a measurement result of 0.

\begin{figure}[t]
  \begin{center}
    \includegraphics[width=\columnwidth]{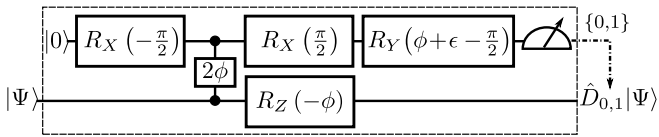}
  \end{center}
  \caption{Quantum circuit using a controlled-phase gate
  \eqref{eq:C-phase} to implement the same partial projections
  $D_{0,1}$ as in Fig.~\ref{fig:circuit}.}
  \label{fig:circuitcphase}
\end{figure}

The realization of the partial projection \eqref{eq:standard} using a
$Z$-controlled $Y$ rotation \eqref{eq:zconty} is shown in
Fig.~\ref{fig:circuit} and Eqs.~\eqref{eq:phi} and (\ref{eq:epsilon}).  Alternatively,
Figure~\ref{fig:circuitcphase} shows how to use a controlled-phase
gate \eqref{eq:C-phase} instead.  In this case, step 2 of the above
procedure is further partitioned into the following steps:
\begin{enumerate}[2a.]
  \item Perform an $X$ rotation of the ancilla by $-\pi/2$.
  \item Perform a controlled-phase entangling gate by an angle $2\phi$.
  \item Perform an $X$ rotation of the ancilla by $\pi/2$.
  \item Perform a $Y$ rotation of the ancilla by $\phi$ to correct
    its phase. (This step can be naturally combined with the step 3 --
    see Fig.~\ref{fig:circuitcphase}.)
  \item Perform a $Z$ rotation of the main qubit by $-\phi$ to correct
    its phase.
\end{enumerate}
This simulation of the $Z$-controlled $Y$ rotation is useful when a
controlled-phase gate is more readily implemented than a direct
$Z$-$Y$ (or $Z$-$X$) coupling interaction.  Note that the final $Z$
rotation of the main qubit may be omitted if the system will be
measured in the $Z$ basis directly after the weak measurement
interaction (e.g., \cite{Groen2013}).

\begin{figure}[t]
  \begin{center}
    \includegraphics[width=\columnwidth]{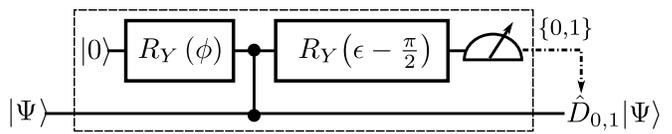}
  \end{center}
  \caption{Quantum circuit using the standard CZ gate
  to implement the same partial projections $D_{0,1}$ as in Fig.\
  \ref{fig:circuit}. The $R_Y(\phi)$ rotation of the ancilla qubit,
  followed by the CZ gate, creates an angular separation of $2\phi$
  between the ancilla states coupled to the qubit states $\ket{0}$
  and $\ket{1}$. Then the ancilla qubit is measured in the slanted
  basis, which becomes the $X$-basis in the symmetric case (when
  $\epsilon=0$).
 }
  \label{fig:circuit2}
\end{figure}

Similarly, Fig.~\ref{fig:circuit2} shows how to replace the
$Z$-controlled $Y$ rotation in the above procedure with a standard
controlled-$Z$ gate $C\!Z(\pi)$.  In this case, step 2 of the
above procedure is instead partitioned into the following steps:
\begin{enumerate}[2a.]
  \item Perform a $Y$ rotation of the ancilla by the angle $\phi$.
  \item Perform a controlled-$Z$ gate to entangle the main qubit with
    the ancilla.
\end{enumerate}
This implementation has the advantage of using a fixed controlled-Z
two-qubit entangling gate, which may be more easily optimized to high
fidelity than $C\!Z(2\phi)$ (e.g., \cite{Barends2013}). As such, the
implementation will be determined by which two-qubit gate has been optimized;
the one-qubit gates typically have high fidelity, even for variable
angles such as $\phi$ and $\epsilon$.

The experimental method discussed in this subsection may suffer from
several types of inefficiency, including imperfect fidelities of the
gates, decoherence during the procedure, and imperfect fidelity of
the ancilla measurement. Nevertheless, we expect this method to give
better overall performance than the thresholded continuous readout
for the current implementations of superconducting qubits.

\section{Generalized Multiple-Outcome Measurements}\label{sec:reduction}

An arbitrary $k$-outcome purity-preserving \cite{purity-pres}
measurement can be implemented by reducing it to a sequence of
two-outcome measurements, which can then be standardized in a
straightforward way.  The resulting decomposition is a sequence of
unitary gates and standardized two-outcome partial projections [Eq.\
\eqref{eq:standard}], each of which can be implemented as discussed
in the previous section.

\subsection{Arbitrary Two-Outcome Measurements}

First, we decompose an arbitrary two-outcome qubit measurement into
the partial projection of Eq.~\eqref{eq:standard} and unitary operations.
Consider a set of two measurement operators $\{N_0,N_1\}$
that correspond to a single two-outcome purity-preserving
measurement \cite{Nielsen2000}. These operators must satisfy the
completeness condition $N_0^\dagger N_0 +
N_1^\dagger N_1 = \openone$. Therefore, the positive
operators $|N_{0,1}| \equiv
(N^\dagger_{0,1}N_{0,1})^{1/2}$ can be diagonalized
simultaneously (with the same unitary operator $V$), and
therefore the singular value decompositions of $N_{0,1}$ have
the form
\begin{align}\label{eq:svd}
  N_{0,1} &= U_{0,1} D_{0,1} V^\dagger,
\end{align}
where $U_{0,1}$ are unitary operators and $D_{0,1}$ are
the (diagonal) partial-projection operators as defined in
Eq.~\eqref{eq:standard}. Notice that Eq.~\eqref{eq:svd} can be
applied to any basis, but we use the natural $Z$-basis, in which the
qubit is (partially) measured.

Therefore, one can implement any two-outcome measurement
$\{N_0,N_1\}$ of a qubit with the following sequence of
operations:
\begin{enumerate}
  \item Apply the unitary operation $V^\dagger$ to the qubit.
  \item Perform the partial projection $D_{0,1}$ using
  specific $p$ and $q$ values [see Eq.~\eqref{eq:standard}].  Record the outcome $0$ or $1$.
  \item Apply the unitary $U_{0}$ or $U_{1}$, depending on the obtained outcome.
\end{enumerate}

\subsection{Reduction Algorithm}
Now we can extend the two-outcome reduction of the previous
subsection to an arbitrary $n$-outcome purity-preserving
measurement. (See also \cite{Andersson2008,Ota2012} for somewhat similar
decompositions.) Consider a set of desired qubit measurement
operators $\{M_0,\cdots,M_{n-1}\}$.  These operators must
satisfy the completeness condition $\sum_{k=0}^{n-1} |M_k|^2 =
\openone$. (Conceptually, one can imagine these operators as
describing the effects on the quantum state that would be induced by
each of the $n$ outcomes of some fictitious laboratory instrument
\cite{Dressel2013b}.)

We can simulate this $n$-outcome measurement by constructing a
sequence of at most $(n-1)$ two-outcome measurements using the
following algorithm:
\begin{enumerate}
  \item Perform the two-outcome measurement with
  $N^{(0)}_0 = M_0$ and $N^{(0)}_1 =
  \sqrt{\openone - |N^{(0)}_0|^2}$.  If $N^{(0)}_1$ is seen, continue
  to the next step.  If $N^{(0)}_0$ is seen, then halt: the net effect on the state is then $N^{(1)}_0 = M_0$.
  \item Perform the two-outcome measurement with
  $N^{(1)}_0 = M_1[N^{(0)}_1]^{-1}$ and
  $N^{(1)}_1 = \sqrt{\openone - |N^{(1)}_0|^2}$.
  If $N^{(1)}_1$ is seen, continue to the next step.  If $N^{(1)}_0$
   is seen, then halt: the net effect on the state is then
  $N^{(1)}_0N^{(0)}_1 = M_1$.
  \item Continue this pattern.  At iteration $k$ measure the outcomes
  $N^{(k)}_0 = M_k[N^{(0)}_1]^{-1} \dots
  [N^{(k-1)}_1]^{-1}$ and $N^{(k)}_1 =
  \sqrt{\openone - |N^{(k)}_0|^2}$.  If $N^{(k)}_1$ is seen, continue
  to step $(k+1)$.  If $N^{(k)}_0$ is seen, then halt: the net effect
  on the state is then $N^{(k)}_0 \dots N^{(1)}_1N^{(0)}_1 =
  M_k$.
  \item Stop at iteration $k=n-2$. A final unitary will in general be necessary
  to make the net effect from last outcome $N^{(n-2)}_1$ match $M_{n-1}$.
\end{enumerate}

Every permutation of the initial set of operators $\{M_k\}$
will produce a different sequence according to this algorithm.  Once
a particular sequence of two-outcome measurements
$\{N^{(k)}_0,N^{(k)}_1\}_{k=0}^{n-2}$ has been constructed
theoretically, each set can then be implemented according to
Eq.~\eqref{eq:svd}.  As a technical improvement, note that at any
intermediate step of this reduction algorithm the operator
$N^{(k)}_1$ can be given an arbitrary unitary degree of
freedom: $N^{(k)}_1 \to U^{(k)}_{\rm add}N^{(k)}_1$.
This extra unitary may be used to eliminate the unitary rotation
$U_1^{(k)}$ from the singular-value decomposition
$N^{(k)}_1$ in Eq.~\eqref{eq:svd}.
 Note that in the described algorithm we implicitly assumed
 $|N^{(k)}_0|^2\leq \openone$; this inequality can be proven in
 a straightforward way.

The described algorithm is general, but it is
not the only possible algorithm for realizing a $n$-outcome
generalized quantum measurement (e.g., \cite{Andersson2008,Ota2012}).
In practice, it will be useful to optimize the algorithm to
produce shorter sequences of measurements.

\section{Fidelity Measures for Generalized Measurement}\label{sec:fidelity}

With the ability to implement generalized quantum measurements comes
the necessity of characterizing how well one is implementing them in
practice.  Currently, there is no standard method for characterizing
fidelity of a generalized measurement with multiple possible
outcomes. There are, however, standard methods for characterizing
the fidelity of individual quantum processes. In this section we
extend these existing definitions to multiple-outcome generalized
measurements (see Appendix for more details). In contrast to the
previous sections, here we assume an arbitrary number $N$ of qubits,
so that the dimension of the Hilbert space is $d=2^N$.

The standard way \cite{Nielsen2000} of describing a quantum
operation $\rho_{\rm in} \mapsto \rho_{\rm fin}$ (where $\rho$
denotes a density matrix) is by using the $d^2\! \times d^2$ process
matrix $\chi$,
\begin{equation}
  \rho \mapsto \sum\nolimits_{i,j} \chi_{ij} E_i \rho E_j^\dagger,
  \label{chi-def}
\end{equation}
where $\{E_i\}$ is the operator basis, which we assume to be the
standard Pauli basis, so that $\Tr{E_j^\dagger E_i}=d \delta_{ij}$.
For a trace-preserving operation, the matrix $\chi$ should satisfy
the condition $\sum_{i,j}\chi_{ij}E^\dagger_j E_i=\openone$, which
in particular implies that $\Tr{\chi}=1$. If the desired quantum
operation $\chi^{\rm ideal}$ corresponds to a \emph{unitary}
rotation, then the fidelity of an experimental
\emph{trace-preserving} operation is usually defined as
\cite{Gilchrist2005} 
\begin{equation}
  F=\Tr{\chi^{\rm ideal}\chi},
  \label{fid-def-usual}
\end{equation}
though sometimes it is defined as the square root of this trace.
 Note that the process fidelity $F$ is linearly related
\cite{Nielsen-02} to the average state fidelity $\overline{F_{\rm
st}}$, which is used in randomized benchmarking and sometimes called
\cite{Magesan-11} ``gate fidelity'', $1-F=(1-\overline{F_{\rm
st}})(1+1/d)$. Also note that the definition (\ref{fid-def-usual})
is inapplicable if two non-unitary operations are compared -- see
the Appendix.

If a desired unitary operation is experimentally realized by a
selective operation (i.e., involving the measurement and selection
of a certain measurement result), then the fidelity definition \eqref{fid-def-usual}
can be modified to \cite{Kiesel-05,Zhong2014} 
\begin{equation}
     F=\frac{\Tr{\chi^{\rm ideal}\chi}}{\Tr{\chi}}.
    \label{fid-def-2}
\end{equation}
In this case, the actual operation is not trace-preserving, so $\Tr{\chi}$
is the selection probability averaged over all initial
pure states (or, equivalently, the selection probability if one prepares
the maximally mixed state). Note that the normalized matrix
$\chi/\Tr{\chi}$ does not correspond to any physical process,
but the definition (\ref{fid-def-2}) still satisfies the requirement
$0\leq F\leq 1$, with $F=1$ only if $\chi /\Tr{\chi} =
\chi^{\rm ideal}$. Also note that other definitions for fidelity of
a selective quantum operation have been considered
\cite{Pedersen-07,Keane-12,Zhong2014}; however, here we will use
Eq.\ (\ref{fid-def-2}) as the starting point for further
generalization.

An experimental realization of a generalized multiple-outcome
quantum measurement can be described as a set of
non-trace-preserving operations, each of them corresponding to a
particular outcome $k$,
\begin{equation}
  \rho \mapsto \sum\nolimits_{i,j}\chi^{(k)}_{ij} E_i \rho E_j^\dagger,
\end{equation}
so that the total nonselective process, $\chi^{\Sigma}=\sum_k
\chi^{(k)}$, is a trace-preserving operation (we assume that at
least one of these outcomes must be reported by the experimental
procedure). The probability of an outcome $k$, averaged over pure
initial states, is
\begin{equation}
p_k=\Tr{\chi^{(k)}},
\end{equation}
such that $\sum_k p_k=1$. For simplicity, we assume that the desired
(ideal) generalized measurement is purity-preserving, $\rho\mapsto
M_k\rho M_k^\dagger$ \cite{Nielsen2000}, where the measurement
operators $M_k$ satisfy the completeness condition $\sum_k
M_k^\dagger M_k=\openone$. Therefore,
\begin{align}
    \chi_{ij}^{(k),\, \rm ideal}&=\alpha^{(k)}_i \alpha^{(k)*}_j, &
    M_k &= \sum\nolimits_{i}\alpha_i^{(k)} E_i,
\end{align}
with $\alpha_i^{(k)}$ being the expansion coefficients of $M_k$ in
the basis $\{E_i\}$.

Each outcome $k$ can then be naturally assigned a ``partial'' fidelity
analogous to Eq.\ (\ref{fid-def-2}) \cite{Kiesel-05,Micuda2008,Bongioanni2010,Zhong2014},
\begin{equation}
F^{(k)}=\frac{\Tr{\chi^{(k),\, \rm ideal}\chi^{(k)}}}{\Tr{\chi^{(k),\, \rm ideal}}\, \Tr{\chi^{(k)}}}.
\label{fid-partial}\end{equation}
Each such fidelity has the proper range, $0\leq F^{(k)}\leq 1$, and is
unity only if $\chi^{(k)} ={\rm const}\times \chi^{(k),\, \rm ideal}$. Note, however,
that this definition is insensitive to the multiplication of $\chi^{(k)}$
by a constant, which affects the average probability $p_k$ of the outcome
$k$. Therefore, to define the overall fidelity $F^{\rm tot}$ of a generalized
measurement as a combination of partial fidelities $F^{(k)}$, we
must ensure that the definition also penalizes for the difference
between the desired probability distribution $p_k^{\rm ideal}=\Tr{\chi^{(k),\rm ideal}}$
and the actual distribution $p_k$.

While there is a significant freedom in defining the overall fidelity $F^{\rm tot}$,
here we suggest two different definitions that in
our opinion are the most natural (see the Appendix for more discussion
on this point). The first definition is
\begin{equation}
 F^{\text{tot}} = \sum_k\,\frac{\Tr{\chi^{(k),\rm ideal}\chi^{(k)}}}
 {\sqrt{\Tr{\chi^{(k),\rm ideal}}\,\Tr{\chi^{(k)}}}}.
    \label{F-tot-1}
\end{equation}
It is obtained as the weighted sum of the partial fidelities (\ref{fid-partial}),
$F^{\rm tot}=\sum_k \sqrt{p_k^{\rm ideal}p_k} \, F^{(k)}$, so that
the weight factors $\sqrt{p_k^{\rm ideal}p_k}$ naturally correspond
to the outcome probabilities and also automatically penalize the
fidelity for unequal probability distributions $p_k^{\rm ideal}$ and
$p_k$.

The second definition we suggest is
\begin{equation}
\tilde F^{\rm tot} = \left[ \sum\nolimits_k\sqrt{\Tr{\chi^{(k),\rm
ideal} \chi^{(k)}}} \right]^2,
    \label{F-tot-2}
\end{equation}
which is obtained from the partial fidelities $F^{(k)}$ as $\tilde F^{\rm
tot}=[\sum_k \sqrt{ p_k^{\rm ideal}p_k \, F^{(k)}}]^2$. The square
root of this definition is a weighted sum of the square roots of the partial
fidelities (\ref{fid-partial}), which in turn are closely related
to the Bhattacharyya coefficient defining the fidelity between classical
probability distributions (see Appendix for more details).

For both definitions (\ref{F-tot-1}) and (\ref{F-tot-2}), the
fidelity is between 0 and 1, and the value of 1 is achieved only if
$\chi^{(k)}=\chi^{(k),\rm ideal}$ for all outcomes $k$. Both
definitions are symmetric under exchange of $\chi^{(k)}
\leftrightarrow \chi^{(k),\text{ideal}}$. Both of them become
inapplicable if the desired generalized measurement includes
decoherence (i.e., does not preserve purity).  (See the Appendix
for the simple generalization that admits decoherence.)

The fidelity definitions (\ref{F-tot-1}) and (\ref{F-tot-2}) compare
an experimentally implemented generalized quantum measurement to an
ideal one, which is characterized by a desired probability
distribution of outcomes (for a given initial state) and their
associated post-measurement states. However, in some experiments the
generalized measurement may be used only to produce desired
probabilities of outcomes, while the post-measurement state is not
important. In this case the fidelity should be defined in a different way.

The probability $P_k(\rho)$ of an outcome $k$ for the
initial state $\rho$ is
\begin{align}
P_k(\rho)&=\Tr{{\cal P}_k\rho }, &
{\cal P}_k&=\sum_{i,j}\chi_{ij}^{(k)}E_j^\dagger E_i,
\end{align}
where ${\cal P}_k$ are so-called POVM elements, which are
positive matrices satisfying the completeness condition $\sum_k
{\cal P}_k=\openone$. Note that the average probability $p_k$
introduced earlier is $p_k=P_k(\openone/d)$. To define a
``probability-only'' fidelity  $F_p$ we need to compare the set of
POVM elements ${\cal P}_k$ with the desired set  ${\cal P}_k^{\rm
ideal}=M_k^\dagger M_k$. Following the same logic as used above for
the process matrices, we define $F_p$ via the weighted sum (with
weights $\sqrt{p_k^{\rm ideal}p_k}$) of the partial fidelities
between (normalized) ${\cal P}_k$ and ${\cal P}_k^{\rm ideal}$, for
which we use either the Uhlmann formula \cite{Nielsen2000} or its
square. This leads to the following two definitions,
    \begin{align}
F_p&= \frac{1}{d}\sum_k \frac{\left({\rm Tr}
 \sqrt{\sqrt{{\cal P}_k^{\rm ideal}} \, {\cal P}_k
 \sqrt{{\cal P}_k^{\rm ideal}}  } \right)^2}
 {\sqrt{\Tr{{\cal P}_k^{\rm ideal}}\, \Tr{{\cal P}_k} }},
    \label{F-POVM-1}\\
\tilde{F}_p&= \left( \frac{1}{d}\sum\nolimits_k {\rm Tr}
 \sqrt{\sqrt{{\cal P}_k^{\rm ideal}} \, {\cal P}_k
 \sqrt{{\cal P}_k^{\rm ideal}}  } \right)^2,
    \label{F-POVM-2}\end{align}
which correspond to the logic of Eqs. (\ref{F-tot-1}) and
(\ref{F-tot-2}), respectively. Note that $\Tr{{\cal
P}_k}=\Tr{\chi^{(k)}}\,d=p_kd$, which produces the factors $d^{-1}$
in the definitions of $F_p$ and $\tilde{F}_p$. It is easier to
determine ${\cal P}_k$ experimentally than $\chi^{(k)}$, because the
matrix ${\cal P}_k$ has dimension $d\times d$, in contrast to
$d^2\!\times d^2$ for $\chi^{(k)}$.

\section{Conclusion}\label{sec:conclusion}

In this paper we have shown that any purity-preserving generalized
measurement of a single qubit can be realized by a combination of
unitary rotations and two-outcome partial-projection measurements.
Two different methods for implementing these partial projections
using superconducting qubits were considered: a thresholded
continuous measurement using a phase-sensitive amplifier, and an
indirect ancilla qubit measurement that uses standard unitary gates
and projective measurements.  The former requires high quantum
efficiency of continuous measurement, while the latter requires
high-fidelity gates and high-fidelity projective measurements. Both
of these methods are already viable experimentally.

The thresholding technique is notable because it realizes a
previously proposed decomposition of generalized measurements into
continuous measurement procedures.  This decomposition is fairly
straightforward in the case of qubits; for higher-dimensional
systems it can also be done, but will generally require more
sophisticated control and feedback of the measurement process.

We have also addressed the issue of characterizing the fidelity of a
generalized quantum measurement with multiple possible outcomes, for
which there is no established definition in
the literature. We proposed two alternative definitions of fidelity
for an experimental generalized measurement, each following slightly 
different logic.

Several special cases of a generalized quantum measurement have
already been realized with superconducting qubits, which essentially
implement both thresholding-based and ancilla-based techniques
similar to those discussed here. We expect that experiments with
generalized measurement will become more routine in the future and
will continue to attract interest, in particular due to potential
practical advantages in applications.

\begin{acknowledgments}
The research was funded by the Office of the Director of National Intelligence (ODNI), Intelligence Advanced Research Projects Activity (IARPA), through the Army Research Office (ARO) Grant No. W911NF-10-1-0334. All statements of fact, opinion, or conclusions contained herein are those of the authors and should not be construed as representing the official views or policies of IARPA, the ODNI, or the U.S. Government. We also acknowledge support from the ARO MURI Grant No. W911NF-11-1-0268.
\end{acknowledgments}

\appendix*
\section{Discussion of Fidelity Measures}

In this Appendix we justify the definitions of fidelity presented in
Sec.~\ref{sec:fidelity} of the main text. Let us start with reviewing existing
definitions of fidelity for probability distributions, density
matrices and quantum processes, and discuss how they relate to one
another.

  Suppose that one experimentally determines a classical probability
distribution $\{p_k\}$ with $\sum_k p_k = 1$ as a set of measured
frequencies. There are several ways to define a
characteristic comparing this distribution to an ideal (reference)
distribution
$\{p^{\text{ideal}}_k\}$. The most widely used characteristics
\cite{Nielsen2000} are the Kolmogorov distance $\sum_k \frac{1}{2}
|p_k-p_k^{\rm ideal}|$ (this is the maximum difference between
probabilities of an event combining several outcomes) and the
\emph{Bhattacharyya coefficient}
\begin{align}\label{eq:classicalfidelity-1}
  F_1(\{p_k\},\{p_k^{\text{ideal}}\}) &= \sum_k \sqrt{p_k\,p_k^{\text{ideal}}},
\end{align}
which is the characteristic that is most relevant to our approach
for defining fidelities.  The Bhattacharyya coefficient has the
intuitive geometric meaning of the cosine of the angle $\theta$
between the two ``probability amplitude'' vectors
$(\sqrt{p_1},\sqrt{p_2},...)$ and
$(\sqrt{p_1^{\text{ideal}}},\sqrt{p_2^{\text{ideal}}},...)$. The
separation angle $\theta$ is also the angle between quantum state
vectors $|\psi\rangle$ and $|\psi^{\rm ideal}\rangle$  that
reproduce these classical probability amplitudes
\cite{Wootters1981}, which is a useful connection.

In spite of nice mathematical properties of the definition
(\ref{eq:classicalfidelity-1}), it has become fashionable in some
quantum computing communities to use this definition {\it squared}
as the fidelity between two probability distributions,
\begin{align}\label{eq:classicalfidelity}
  F_2(\{p_k\},\{p_k^{\text{ideal}}\}) &= \left[\sum_k \sqrt{p_k\,p_k^{\text{ideal}}}\right]^2.
\end{align}
This change in definition is primarily because the squared definition
has a direct connection to the standard overlap $|\langle\psi | \psi^{\rm
ideal}\rangle |^2$ between two wavefunctions, which in turn is
related to the probability of a quantum measurement result when 
$|\psi\rangle$ is measured ``along'' $| \psi^{\rm ideal}\rangle$.

The choice between the two definitions
(\ref{eq:classicalfidelity-1}) and (\ref{eq:classicalfidelity}) has
essentially doubled the number of fidelity definitions that are used
in quantum computing, which has created some confusion. As discussed
later, our proposed fidelity definitions (\ref{F-tot-1}) and
(\ref{F-POVM-1}) follow the logic of the definition
{\eqref{eq:classicalfidelity} [while using
\eqref{eq:classicalfidelity-1} for the weight factors]. In contrast,
our definitions (\ref{F-tot-2}) and (\ref{F-POVM-2}) follow the
logic of Eq.\ (\ref{eq:classicalfidelity-1}), but at the end are
converted (by squaring) into the more standard ``dimension'' of the
definition  (\ref{eq:classicalfidelity}).

Now suppose that one experimentally determines a quantum state
$\rho$ with $\Tr{\rho} = 1$ using quantum state tomography). The
Uhlmann fidelity \cite{Uhlmann1976,Nielsen2000} of this state
compared to a reference (ideal) state $\rho^{\text{ideal}}$ is
usually defined as
\begin{equation}
  F_3(\rho,\rho^{\text{ideal}}) = \Tr{\sqrt{\sqrt{\rho^{\text{ideal}}}\,
  \rho\,\sqrt{\rho^{\text{ideal}}}}},
    \label{Uhlmann-1}
\end{equation}
but it can also be defined \cite{Jozsa-94,Gilchrist2005} as its square,
\begin{equation}
      F_4(\rho,\rho^{\text{ideal}})= \left[
      F_3(\rho,\rho^{\text{ideal}})\right]^2 .
    \label{Uhlmann-squared}
\end{equation}
Importantly, the Uhlmann fidelity (\ref{Uhlmann-1}) can be found
by minimizing the Bhattacharyya coefficient \eqref{eq:classicalfidelity-1}
over all possible generalized measurements that can be made on the
two states to produce probability distributions to be compared
\cite{Barnum-96},
\begin{equation}
  F_3(\rho,\rho^{\text{ideal}}) = \min_{\{{\cal P}_k\}}
  F_1\left[\{\Tr{{\cal P}_k\rho}\},\,\{\Tr{{\cal P}_k
  \rho^{\text{ideal}}}\}\right],
\end{equation}
where $\{{\cal P}_k\}$ are varied over all possible sets of positive
probability operators such that $\sum_k {\cal P}_k = \openone$.
Squaring this equation produces the same relation between
definitions (\ref{Uhlmann-squared}) and
(\ref{eq:classicalfidelity}).  In most cases the reference state
$\rho^{\text{ideal}} = \pr{\psi^{\rm ideal}}$ is a pure state, in
which case the definition (\ref{Uhlmann-squared}) reduces to the
simpler state overlap,
\begin{align}\label{eq:quantumfidelity}
  F_4 \to F_5(\rho,\rho^{\text{ideal}}) &= \Tr{\rho\,\rho^{\text{ideal}}} =
  \bra{\psi^{\rm ideal}}\rho\ket{\psi^{\rm ideal}}.
\end{align}
As mentioned above, the simplicity of this overlap motivates the
choice of the squared definitions (\ref{Uhlmann-squared}) and
(\ref{eq:classicalfidelity}), which we have adopted here and in
the main text.  (Note that the full Uhlmann formula for $F_4$
does not permit any simple interpretation.)

Now suppose that one experimentally determines a quantum process
matrix $\chi$ with quantum process tomography (QPT).  This matrix is
mathematically equivalent to a (generally unnormalized) density
operator, so the definition of its fidelity compared to an ideal
process matrix $\chi^{\text{ideal}}$ can be based on the fidelity
definition \eqref{Uhlmann-squared} for density matrices. To see
this, recall that for $N$ qubits such a matrix $\chi$ is found by
first choosing a matrix basis $\{ E_i \}$ that usually consists of
all $4^N$ tensor products of the four Pauli operators
$\{I,\sigma_x,\sigma_y,\sigma_z\}$, and then writing the process as
a state-transformation function of the form
\begin{align}
  \rho \mapsto \sum_{i,j} \chi_{i,j}\, E_i\, \rho\, E_j^\dagger,
\end{align}
where $\chi_{i,j}$ are the complex components of the $4^N \times
4^N$ Hermitian process matrix $\chi$.  Typically, the
reference (ideal) process is assumed to be purity-preserving, and
thus characterized by a single Kraus operator $M$: $\rho \mapsto M \rho
M^\dagger$.  This operator can be expanded in terms of the Pauli
matrix basis as $M = \sum_i \alpha_i \, E_i$ where $\alpha_i =
\Tr{E_i^\dagger M}/2^N$ are its complex components.  Hence, the
reference process matrix components have the form
\begin{align}
  \chi^{\text{ideal}}_{i,j} &= \alpha_i \alpha_j^*.
\end{align}
This expression can be related to a density matrix by formally
defining a complex vector $\ket{M}$ of the components $\alpha_i$ of
$M$ and then expressing the reference process matrix
$\chi^{\text{ideal}}$ as a dyadic (outer) product
\begin{align}
  \ket{M} &= (\alpha_1,\cdots,\alpha_{4^N})^T, \\
  \chi^{\text{ideal}} &= \pr{M}.
\label{chi-via-M}\end{align}

If $M$ is unitary then the reference process is trace-preserving and
$\Tr{\chi^{\text{ideal}}} = 1$, so $\chi^{\text{ideal}}$ is
completely equivalent to a pure state density operator. If the experimental
process $\chi$ is also trace preserving, then it is also equivalent to a
density operator acting in the Hilbert space with dimension
$2^{2N}$, corresponding to a generally mixed state (the
Jamio{\l}kowski-Choi ``channel-state duality''
\cite{Jamiolkowski1974,Choi1975,Min2013}). Therefore, the fidelity
definition (\ref{eq:quantumfidelity}) can be used directly, leading
to the standard definition  \cite{Gilchrist2005,Raginsky2001}
\begin{equation}
    F_6(\chi,\chi^{\text{ideal}}) = \Tr{\chi\,\chi^{\text{ideal}}}
    \label{fid-chi-st}
\end{equation}
for the fidelity of a quantum process $\chi$. If the reference
process $\chi^{\rm ideal}$ is not unitary (but is still
trace-preserving), then this definition is naturally replaced with the
definition based on Eq.\ (\ref{Uhlmann-squared}),
\begin{equation}
     F_7(\chi,\chi^{\text{ideal}}) = \left[ \Tr{
\sqrt{ \sqrt{\chi^{\rm ideal}} \, \chi \sqrt{\chi^{\rm ideal}}}
  }\right]^2.
    \label{fid-chi-st-2}
\end{equation}
Choosing the non-squared fidelity definition $(\ref{Uhlmann-1})$
instead will produce the equally valid process fidelity definition
$\sqrt{F_7}$ (and correspondingly $\sqrt{F_6}$); however, this
variation is not typically used in QPT experiments, so we do not
consider it here.

If the process $\chi$ is not trace-preserving, it necessarily
involves a selection; in other words, we consider the process as
happening only in some ``successful'' cases (e.g., when  a detector
clicks). There are several meaningful ways to generalize the
definitions (\ref{fid-chi-st}) and (\ref{fid-chi-st-2}) to this
case. For example, if the ideal process is still unitary, we can
continue using the standard definition $F_6$ (\ref{fid-chi-st})
without any change. This will mean that we take into account all
realizations of the process, including ``unsuccessful'' ones, for
which we assign zero fidelity. Alternatively, we can consider only
``successful'' realizations. In this case there are also several
ways to generalize the standard fidelity definition (e.g.\
\cite{Kor-Keane-10}); here we will consider the way that is based
on the Jamio{\l}kowski-Choi channel-state duality.

For non-unitary $M$ in Eq.\ (\ref{chi-via-M}) (e.g., a partial
projection), dividing $\chi^{\text{ideal}}$ by its trace
$\Tr{\chi^{\text{ideal}}} = \braket{M}{M} = \sum_i |\alpha_i|^2$
still produces a positive matrix with unit trace that is formally
equivalent to a pure state density operator. Similarly,
$\chi/\Tr{\chi}$ is a positive matrix with unit trace, and therefore
it is also formally equivalent to a (generally mixed) density
operator. As a result, the definition \cite{Kiesel-05,Micuda2008,Bongioanni2010,Zhong2014}
\begin{align}\label{eq:processfidelity}
  F_8(\chi,\chi^{\text{ideal}}) = \frac{\Tr{\chi\,\chi^{\text{ideal}}}}
 {\Tr{\chi}\,\Tr{\chi^{\text{ideal}}}} =
\frac{\bra{M}\chi\ket{M}}{\Tr{\chi}\,\braket{M}{M}}.
\end{align}
satisfies the condition $0\leq F_8\leq 1$, with $F_8=1$ only if
$\chi = {\rm const} \times \chi^{\rm ideal}$. In the case when both
$\chi$ and $\chi^{\rm ideal}$ are not purity-preserving (for
example, when decoherence is considered even for the ``ideal''
process), both equivalent density operators will be mixed, so it
will be necessary to use the full Uhlmann form of the state fidelity
definition (\ref{Uhlmann-squared}) instead of its simplified form
(\ref{eq:quantumfidelity}) for the corresponding density matrices;
this trivially generalizes Eq.\ (\ref{eq:processfidelity}) to
\begin{equation}\label{eq:fid-chi-gen}
  F_9(\chi,\chi^{\text{ideal}}) = \frac{\left[\Tr{
\sqrt{ \sqrt{\chi^{\rm ideal}} \, \chi \sqrt{\chi^{\rm ideal}}}
  }\right]^2}
 {\Tr{\chi}\,\Tr{\chi^{\text{ideal}}}}.
\end{equation}
We emphasize that the definitions (\ref{eq:processfidelity}) and
(\ref{eq:fid-chi-gen}) compare the two operations only when they are
successfully realized (selected).

\vspace{0.3cm}

Given the fidelity definitions reviewed above, we now consider a
generalized measurement that has several outcomes $k$. Each
distinguishable outcome corresponds to a separate quantum process
\begin{align}
  \rho \mapsto \sum_{i,j}\chi^{(k)}_{i,j}\,E_i\,\rho\,E_j^\dagger,
\end{align}
characterized by a process matrix $\chi^{(k)}$ that can be
determined experimentally with QPT.  The sum of these process
matrices produces the total \emph{nonselective} process matrix that
sums over all possible outcomes: $\sum_k \chi^{(k)} = \chi^{\rm
ns}$. The nonselective process $\chi^{\rm ns}$ will be
trace-preserving (assuming no loss) so the trace of this matrix is
unity.  In contrast, the trace of each outcome matrix is the
probability $p_k = \Tr{\chi^{(k)}}$ of obtaining the outcome $k$ if
one prepares a maximally mixed state (or, equivalently, if one
averages over all possible preparations). All the outcome matrices
$\chi^{(k)}$ and their associated probabilities $p_k$ should be
involved in the definition of the total fidelity of the generalized
measurement.

The reference measurement will typically have purity-preserving
processes $\chi^{(k),\rm ideal}$ for all outcomes $k$ that are
completely characterized by single Kraus operators $M^{(k)}$, as
discussed above, along with their associated component vectors
$\ket{M^{(k)}}$, process matrices $\chi^{(k),\text{ideal}} =
\pr{M^{(k)}}$, and outcome probabilities $p_k^{\text{ideal}} =
\Tr{\chi^{(k),\text{ideal}}} = \braket{M^{(k)}}{M^{(k)}}$.  The
total nonselective process matrix $\chi^{\text{ns,ideal}}$ will 
also typically be trace-preserving (and thus have unit trace).

Defining a sensible overall fidelity $F^{\text{tot}}$ that properly
includes information about all outcomes $k$ involves the following
basic conceptual constraints:
\begin{enumerate}[(a)]
  \item $F^{\text{tot}}$ should be a symmetric function of all the
matrices $\chi^{(k)}$, so the outcomes are interchangeable.
  \item The definition $F^{\text{tot}}$ should be in the range $[0,1]$.
  \item $F^{\text{tot}} = 0$ only when $F_{9}(\chi^{(k)},\chi^{(k),
\text{ideal}}) = 0$ for all $k$. (Note that $F_9$ reduces to $F_8$
for purity-preserving ideal processes. This applies to all
discussions below.)
  \item $F^{\text{tot}} = 1$ only when $F_{9}(\chi^{(k)},\chi^{(k),
\text{ideal}}) = 1$ for all $k$ and
$F_2(\{p_k\},\{p_k^{\text{ideal}}\}) = 1$. (This implies
    $p_k = p_k^{\text{ideal}}$ for all $k$, and therefore $\chi^{(k)}=
    \chi^{(k),\text{ideal}}$).
  \item The definition should be symmetric under the exchange of
    $\chi^{(k)} \leftrightarrow \chi^{(k),\text{ideal}}$.
\end{enumerate}
To satisfy these constraints, candidate definitions should be
constructed from other meaningful quantities in the range $[0,1]$,
such as the symmetric classical fidelities
$F_{2}(\{p_k\},\{p_k^{\text{ideal}}\})$ (or alternatively $F_1$),
the outcome process fidelities
$F_{9}(\chi^{(k)},\chi^{(k),\text{ideal}})$, and the outcome
probabilities $p_k,p_k^{\text{ideal}}$, which are all functions of
the matrices $\chi^{(k)}$.

A simple choice for a candidate definition that satisfies
all the above constraints and includes each outcome probability and
fidelity explicitly is
\begin{align}\label{eq:functionalform}
  F^{\text{tot}} &= C\times \left[\sum_k\,\sqrt{p_k p_k^{\text{ideal}}}\,
  F_{9}(\chi^{(k)},\chi^{(k),\text{ideal}})^\alpha\right]^\beta,
\end{align}
where $\alpha,\beta$ are real numbers and the factor $C$ is
discussed below. The symmetric weighting factors $\sqrt{p_k p_k^{\rm
ideal}}$ automatically penalize for unequal outcome probability
distributions $p_k$ and $p_k^{\rm ideal}$, while the symmetric
fidelities $F_{9}^\alpha$ penalize for the differences between each
outcome separately. The power $\alpha$ determines the relative
importance of these penalties, while $\beta$ is the overall power.
The optional prefactor $C$ can contain any number of additional
penalization factors that independently satisfy the above
constraints.  Examples of factors that can be included in $C$ are:
\begin{itemize}
  \item $C_1 = \left[\sum_k \sqrt{p_k p_k^{\text{ideal}}}\right]^{\beta_1},$
  \item $C_2 = \left[\frac{1}{n}\sum_k F_9(\chi^{(k)},\chi^{(k),\text{ideal}})^{\alpha_2}\right]^{\beta_2},$
  \item $C_3 = \left[\sum_k p_k F_9(\chi^{(k)},\chi^{(k),\text{ideal}})^{\alpha_3}\right]^{\beta_3},$
  \item $C_4 = \left[\sum_k p_k^{\text{ideal}} F_9(\chi^{(k)},\chi^{(k),\text{ideal}})^{\alpha_4}\right]^{\beta_4},$
  \item $C_5 = \prod_k F_9(\chi^{(k)},\chi^{(k),\text{ideal}})^{\alpha^{(k)}_5}$,
\end{itemize}
where $n$ is the number of outcomes, each $\alpha_{2,3,4}$ and $\beta_{1,2,3,4}$ 
are real numbers, and the outcome-dependent weights $\alpha^{(k)}_5$ can be 
chosen as $\alpha^{(k)}_5 = p_k$,  or $p_k^{\text{ideal}}$.
Note that the examples $C_3$, $C_4$ and $C_5$ break the symmetry
constraint unless they are properly combined. Also note that $C_2$, $C_3$, 
$C_4$, and $C_5$ do not penalize for $p_k \neq p_k^{\rm ideal}$, so 
they can replace the main term in Eq.\ (\ref{eq:functionalform})
only if additionally multiplied by $C_1$. While each of these
factors can penalize the total fidelity in interesting ways, we
choose the simplest functional form with $C = 1$ as the most
practical definition.

The remaining parameters $\alpha,\beta$ in Eq.\
(\ref{eq:functionalform}) can be constrained by requiring $F^{\rm
tot}$ to consistently reduce to the existing definitions of fidelity
as limiting cases. To match the form of the single-outcome fidelity
$F_{7}$ and/or $F_6$ when only one $p_k=p_k^{\rm ideal}=1$ with the
rest zero, we need to choose $\beta=1/\alpha$, so we are left with
only one free parameter $\alpha$.  This parameter can be chosen by
matching with the classical probability fidelities $F_{1}$ or $F_2$
when $F_{9}(\chi^{(k)},\chi^{(k),\text{ideal}})=1$ for all $k$; this
gives $\alpha=1$ ($\beta=1$) or $\alpha = 1/2$ ($\beta=2$),
correspondingly.

The choice of $\alpha = 1$ identifies $F_1$ as the preferred
classical fidelity for the sum over $k$, yielding the
definition
\begin{align}\label{eq:measurementfidelity1}
F^{\text{tot}}_1 &= \sum_k\,\sqrt{p_k p_k^{\text{ideal}}}\,
F_{9}(\chi^{(k)},\chi^{(k),\text{ideal}}),
\end{align}
which reduces to Eq.~\eqref{F-tot-1} in the main text when the
simplified form $F_8$ is used for the individual outcome fidelities,
\begin{equation}
   F^{\rm tot}_1 \to \sum_k\,\frac{\Tr{\chi^{(k)}\,\chi^{(k),\text{ideal}}}}
{\sqrt{\Tr{\chi^{(k)}}\,\Tr{\chi^{(k),\text{ideal}}}}}.
\label{fid-def1-repeat}\end{equation}

The alternative choice of $\alpha = 1/2$ consistently identifies
$F_2$ as the preferred classical fidelity for both the sum over $k$
and each individual outcome fidelities $F_{9}$. This choice produces
the definition
\begin{align}\label{eq:measurementfidelity2}
  F^{\text{tot}}_2 = \left[\sum_k (p_k p_k^{\text{ideal}})^{1/2} \,
F_{9}(\chi^{(k)},\chi^{(k),\text{ideal}})^{1/2}\right]^2,
\end{align}
which reduces to the definition \eqref{F-tot-2} in the main text
when the simplified form $F_8$ is used,
\begin{align}
  F^{\text{tot}}_2 \to \left[\sum_k \sqrt{\Tr{\chi^{(k)}\chi^{(k),\text{ideal}}}}
\right]^2.
  \label{fid-def2-repeat}\end{align}

The advantage of the definition (\ref{fid-def1-repeat}) is that it
is a linear combination of the outcome fidelities $F_{8}$. However,
because of the denominator, it is a complicated function of the
process matrices.  This complication is related to a subtle
inconsistency in the definition: while $F_1$ is chosen as the
preferred classical fidelity for the sum over $k$, $F_2$ has been
chosen as the preferred classical fidelity that matches the
squared-form of the fidelity $F_{8}$.  Thus, the compromise between
the two choices of classical fidelity preserves the linearity of
$F^{\rm tot}_1$ in the outcome fidelities $F_{8}$, but makes the
final expression in terms of process matrices complicated.

Since this definition (\ref{fid-def2-repeat}) is logically
consistent in choosing $F_2$ for both cases, its form in terms of
process matrices is simpler.  In fact, the only difference between
this definition and the classical fidelity in
\eqref{eq:classicalfidelity} is that the factor inside the square
root is an overlap of noncommutative $\chi_k$ matrices, rather than
the product of classical probabilities. Furthermore, removing the
outer square from this definition consistently chooses $F_1$ as the
preferred classical fidelity for both the sum over $k$ and each
(non-squared) outcome fidelity $\sqrt{F_{9}}$. As a result, the
total fidelity $\sqrt{F^{\rm tot}_2}$ becomes a linear function of
the non-squared outcome fidelities $\sqrt{F_{9}}$ automatically.

As noted earlier, more complicated definitions of $F^{\rm tot}$ can
also be considered. However, we feel that the two simplest
definitions \eqref{eq:measurementfidelity1} and
\eqref{eq:measurementfidelity2} will be the most useful in practice.
The ``probability-only'' definitions (\ref{F-POVM-1}) and
(\ref{F-POVM-2}) in the main text are also essentially the
definitions \eqref{eq:measurementfidelity1} and
\eqref{eq:measurementfidelity2}, but applied to the POVM elements
${\cal P}_k$ instead of the process matrices $\chi^{(k)}$ [the extra
factor $d^{-1}$ comes from different normalizations: $\Tr{{\cal
P}_k}=\Tr{ \chi^{(k)}}\,d$].

%\bibliography{citations}

\begin{thebibliography}{90}%
\makeatletter
\providecommand \@ifxundefined [1]{%
 \@ifx{#1\undefined}
}%
\providecommand \@ifnum [1]{%
 \ifnum #1\expandafter \@firstoftwo
 \else \expandafter \@secondoftwo
 \fi
}%
\providecommand \@ifx [1]{%
 \ifx #1\expandafter \@firstoftwo
 \else \expandafter \@secondoftwo
 \fi
}%
\providecommand \natexlab [1]{#1}%
\providecommand \enquote  [1]{``#1''}%
\providecommand \bibnamefont  [1]{#1}%
\providecommand \bibfnamefont [1]{#1}%
\providecommand \citenamefont [1]{#1}%
\providecommand \href@noop [0]{\@secondoftwo}%
\providecommand \href [0]{\begingroup \@sanitize@url \@href}%
\providecommand \@href[1]{\@@startlink{#1}\@@href}%
\providecommand \@@href[1]{\endgroup#1\@@endlink}%
\providecommand \@sanitize@url [0]{\catcode `\\12\catcode `\$12\catcode
  `\&12\catcode `\#12\catcode `\^12\catcode `\_12\catcode `\%12\relax}%
\providecommand \@@startlink[1]{}%
\providecommand \@@endlink[0]{}%
\providecommand \url  [0]{\begingroup\@sanitize@url \@url }%
\providecommand \@url [1]{\endgroup\@href {#1}{\urlprefix }}%
\providecommand \urlprefix  [0]{URL }%
\providecommand \Eprint [0]{\href }%
\providecommand \doibase [0]{http://dx.doi.org/}%
\providecommand \selectlanguage [0]{\@gobble}%
\providecommand \bibinfo  [0]{\@secondoftwo}%
\providecommand \bibfield  [0]{\@secondoftwo}%
\providecommand \translation [1]{[#1]}%
\providecommand \BibitemOpen [0]{}%
\providecommand \bibitemStop [0]{}%
\providecommand \bibitemNoStop [0]{.\EOS\space}%
\providecommand \EOS [0]{\spacefactor3000\relax}%
\providecommand \BibitemShut  [1]{\csname bibitem#1\endcsname}%
\let\auto@bib@innerbib\@empty
%</preamble>
\bibitem [{\citenamefont {Dirac}(1930)}]{Dirac1930}%
  \BibitemOpen
  \bibfield  {author} {\bibinfo {author} {\bibfnamefont {P.~A.~M.}\
  \bibnamefont {Dirac}},\ }\href@noop {} {\emph {\bibinfo {title} {{Principles
  of Quantum Mechanics}}}}\ (\bibinfo  {publisher} {Oxford University Press,
  Oxford},\ \bibinfo {year} {1930})\BibitemShut {NoStop}%
\bibitem [{\citenamefont {von Neumann}(1932)}]{VonNeumann1932}%
  \BibitemOpen
  \bibfield  {author} {\bibinfo {author} {\bibfnamefont {J.}~\bibnamefont {von
  Neumann}},\ }\href@noop {} {\emph {\bibinfo {title} {Mathematische Grundlagen
  der Quantenmechanik}}}\ (\bibinfo  {publisher} {Berlin: Springer},\ \bibinfo
  {year} {1932})\BibitemShut {NoStop}%
\bibitem [{\citenamefont {Jauch}(1968)}]{Jauch1968}%
  \BibitemOpen
  \bibfield  {author} {\bibinfo {author} {\bibfnamefont {J.~M.}\ \bibnamefont
  {Jauch}},\ }\href@noop {} {\emph {\bibinfo {title} {Foundations of quantum
  mechanics}}}\ (\bibinfo  {publisher} {Addison-Wesley Pub. Co., Reading},\
  \bibinfo {year} {1968})\BibitemShut {NoStop}%
\bibitem [{\citenamefont {Davies}(1976)}]{Davies1976}%
  \BibitemOpen
  \bibfield  {author} {\bibinfo {author} {\bibfnamefont {E.~B.}\ \bibnamefont
  {Davies}},\ }\href@noop {} {\emph {\bibinfo {title} {{Quantum Theory of Open
  Systems}}}}\ (\bibinfo  {publisher} {Academic, London},\ \bibinfo {year}
  {1976})\BibitemShut {NoStop}%
\bibitem [{\citenamefont {Holevo}(1982)}]{Holevo1982}%
  \BibitemOpen
  \bibfield  {author} {\bibinfo {author} {\bibfnamefont {A.~S.}\ \bibnamefont
  {Holevo}},\ }\href@noop {} {\emph {\bibinfo {title} {{Probabilistic and
  Statistical Aspects of the Quantum Theory}}}}\ (\bibinfo  {publisher}
  {North-Holland, Amsterdam},\ \bibinfo {year} {1982})\BibitemShut {NoStop}%
\bibitem [{\citenamefont {Kraus}(1983)}]{Kraus1983}%
  \BibitemOpen
  \bibfield  {author} {\bibinfo {author} {\bibfnamefont {K.}~\bibnamefont
  {Kraus}},\ }\href@noop {} {\emph {\bibinfo {title} {States, effects and
  operations: fundamental notions of quantum theory}}}\ (\bibinfo  {publisher}
  {Springer-Verlag, Berlin},\ \bibinfo {year} {1983})\BibitemShut {NoStop}%
\bibitem [{\citenamefont {Aharonov}\ \emph {et~al.}(1988)\citenamefont
  {Aharonov}, \citenamefont {Albert},\ and\ \citenamefont
  {Vaidman}}]{Aharonov1988}%
  \BibitemOpen
  \bibfield  {author} {\bibinfo {author} {\bibfnamefont {Y.}~\bibnamefont
  {Aharonov}}, \bibinfo {author} {\bibfnamefont {D.~Z.}\ \bibnamefont
  {Albert}}, \ and\ \bibinfo {author} {\bibfnamefont {L.}~\bibnamefont
  {Vaidman}},\ }\href@noop {} {\bibfield  {journal} {\bibinfo  {journal} {Phys.
  Rev. Lett.}\ }\textbf {\bibinfo {volume} {60}},\ \bibinfo {pages} {1351}
  (\bibinfo {year} {1988})}\BibitemShut {NoStop}%
\bibitem [{\citenamefont {Braginski}\ and\ \citenamefont
  {Khalili}(1992)}]{Braginski1992}%
  \BibitemOpen
  \bibfield  {author} {\bibinfo {author} {\bibfnamefont {V.}~\bibnamefont
  {Braginski}}\ and\ \bibinfo {author} {\bibfnamefont {F.}~\bibnamefont
  {Khalili}},\ }\href@noop {} {\emph {\bibinfo {title} {Quantum Measurement}}}\
  (\bibinfo  {publisher} {Cambridge University Press, Cambridge},\ \bibinfo
  {year} {1992})\BibitemShut {NoStop}%
\bibitem [{\citenamefont {Nielsen}\ and\ \citenamefont
  {Chuang}(2000)}]{Nielsen2000}%
  \BibitemOpen
  \bibfield  {author} {\bibinfo {author} {\bibfnamefont {M.~A.}\ \bibnamefont
  {Nielsen}}\ and\ \bibinfo {author} {\bibfnamefont {I.~L.}\ \bibnamefont
  {Chuang}},\ }\href@noop {} {\emph {\bibinfo {title} {Quantum computation and
  quantum information}}}\ (\bibinfo  {publisher} {Cambridge University Press,
  Cambridge},\ \bibinfo {year} {2000})\BibitemShut {NoStop}%
\bibitem [{\citenamefont {Keyl}(2002)}]{Keyl2002}%
  \BibitemOpen
  \bibfield  {author} {\bibinfo {author} {\bibfnamefont {M.}~\bibnamefont
  {Keyl}},\ }\href@noop {} {\bibfield  {journal} {\bibinfo  {journal} {Physics
  Reports}\ }\textbf {\bibinfo {volume} {369}},\ \bibinfo {pages} {431}
  (\bibinfo {year} {2002})}\BibitemShut {NoStop}%
\bibitem [{\citenamefont {Wiseman}\ and\ \citenamefont
  {Milburn}(2009)}]{Wiseman2009}%
  \BibitemOpen
  \bibfield  {author} {\bibinfo {author} {\bibfnamefont {H.~M.}\ \bibnamefont
  {Wiseman}}\ and\ \bibinfo {author} {\bibfnamefont {G.}~\bibnamefont
  {Milburn}},\ }\href@noop {} {\emph {\bibinfo {title} {Quantum Measurement and
  Control}}}\ (\bibinfo  {publisher} {Cambridge University Press, Cambridge},\
  \bibinfo {year} {2009})\BibitemShut {NoStop}%
\bibitem [{\citenamefont {Korotkov}(1991)}]{Korotkov-01}%
  \BibitemOpen
  \bibfield  {author} {\bibinfo {author} {\bibfnamefont {A.~N.}\ \bibnamefont
  {Korotkov}},\ }\href@noop {} {\bibfield  {journal} {\bibinfo  {journal}
  {Phys. Rev. B}\ }\textbf {\bibinfo {volume} {63}},\ \bibinfo {pages} {115403}
  (\bibinfo {year} {1991})}\BibitemShut {NoStop}%
\bibitem [{\citenamefont {Dressel}\ and\ \citenamefont
  {Jordan}(2012)}]{Dressel2012b}%
  \BibitemOpen
  \bibfield  {author} {\bibinfo {author} {\bibfnamefont {J.}~\bibnamefont
  {Dressel}}\ and\ \bibinfo {author} {\bibfnamefont {A.~N.}\ \bibnamefont
  {Jordan}},\ }\href@noop {} {\bibfield  {journal} {\bibinfo  {journal} {Phys.
  Rev. A}\ }\textbf {\bibinfo {volume} {85}},\ \bibinfo {pages} {022123}
  (\bibinfo {year} {2012})}\BibitemShut {NoStop}%
\bibitem [{\citenamefont {Huttner}\ \emph {et~al.}(1996)\citenamefont
  {Huttner}, \citenamefont {Muller}, \citenamefont {Gautier}, \citenamefont
  {Zbinden},\ and\ \citenamefont {Gisin}}]{Huttner1996}%
  \BibitemOpen
  \bibfield  {author} {\bibinfo {author} {\bibfnamefont {B.}~\bibnamefont
  {Huttner}}, \bibinfo {author} {\bibfnamefont {A.}~\bibnamefont {Muller}},
  \bibinfo {author} {\bibfnamefont {J.~D.}\ \bibnamefont {Gautier}}, \bibinfo
  {author} {\bibfnamefont {H.}~\bibnamefont {Zbinden}}, \ and\ \bibinfo
  {author} {\bibfnamefont {N.}~\bibnamefont {Gisin}},\ }\href@noop {}
  {\bibfield  {journal} {\bibinfo  {journal} {Phys. Rev. A}\ }\textbf {\bibinfo
  {volume} {54}},\ \bibinfo {pages} {3783} (\bibinfo {year}
  {1996})}\BibitemShut {NoStop}%
\bibitem [{\citenamefont {Pryde}\ \emph {et~al.}(2004)\citenamefont {Pryde},
  \citenamefont {O'Brien}, \citenamefont {White}, \citenamefont {Bartlett},\
  and\ \citenamefont {Ralph}}]{Pryde2004}%
  \BibitemOpen
  \bibfield  {author} {\bibinfo {author} {\bibfnamefont {G.~J.}\ \bibnamefont
  {Pryde}}, \bibinfo {author} {\bibfnamefont {J.~L.}\ \bibnamefont {O'Brien}},
  \bibinfo {author} {\bibfnamefont {A.}~\bibnamefont {White}}, \bibinfo
  {author} {\bibfnamefont {S.}~\bibnamefont {Bartlett}}, \ and\ \bibinfo
  {author} {\bibfnamefont {T.}~\bibnamefont {Ralph}},\ }\href@noop {}
  {\bibfield  {journal} {\bibinfo  {journal} {Phys. Rev. Lett.}\ }\textbf
  {\bibinfo {volume} {92}},\ \bibinfo {pages} {190402} (\bibinfo {year}
  {2004})}\BibitemShut {NoStop}%
\bibitem [{\citenamefont {Guerlin}\ \emph {et~al.}(2007)\citenamefont
  {Guerlin}, \citenamefont {Bernu}, \citenamefont {Del\'eglise}, \citenamefont
  {Sayrin}, \citenamefont {Gleyzes}, \citenamefont {Kuhr}, \citenamefont
  {Brune}, \citenamefont {Raimond},\ and\ \citenamefont
  {Haroche}}]{Guerlin2007}%
  \BibitemOpen
  \bibfield  {author} {\bibinfo {author} {\bibfnamefont {C.}~\bibnamefont
  {Guerlin}}, \bibinfo {author} {\bibfnamefont {J.}~\bibnamefont {Bernu}},
  \bibinfo {author} {\bibfnamefont {S.}~\bibnamefont {Del\'eglise}}, \bibinfo
  {author} {\bibfnamefont {C.}~\bibnamefont {Sayrin}}, \bibinfo {author}
  {\bibfnamefont {S.}~\bibnamefont {Gleyzes}}, \bibinfo {author} {\bibfnamefont
  {S.}~\bibnamefont {Kuhr}}, \bibinfo {author} {\bibfnamefont {M.}~\bibnamefont
  {Brune}}, \bibinfo {author} {\bibfnamefont {J.-M.}\ \bibnamefont {Raimond}},
  \ and\ \bibinfo {author} {\bibfnamefont {S.}~\bibnamefont {Haroche}},\
  }\href@noop {} {\bibfield  {journal} {\bibinfo  {journal} {Nature}\ }\textbf
  {\bibinfo {volume} {448}},\ \bibinfo {pages} {889} (\bibinfo {year}
  {2007})}\BibitemShut {NoStop}%
\bibitem [{\citenamefont {Gleyzes}\ \emph {et~al.}(2007)\citenamefont
  {Gleyzes}, \citenamefont {Kuhr}, \citenamefont {Guerlin}, \citenamefont
  {Bernu}, \citenamefont {Del\'eglise}, \citenamefont {Hoff}, \citenamefont
  {Brune}, \citenamefont {Raimond},\ and\ \citenamefont
  {Haroche}}]{Gleyzes2007}%
  \BibitemOpen
  \bibfield  {author} {\bibinfo {author} {\bibfnamefont {S.}~\bibnamefont
  {Gleyzes}}, \bibinfo {author} {\bibfnamefont {S.}~\bibnamefont {Kuhr}},
  \bibinfo {author} {\bibfnamefont {C.}~\bibnamefont {Guerlin}}, \bibinfo
  {author} {\bibfnamefont {J.}~\bibnamefont {Bernu}}, \bibinfo {author}
  {\bibfnamefont {S.}~\bibnamefont {Del\'eglise}}, \bibinfo {author}
  {\bibfnamefont {U.~B.}\ \bibnamefont {Hoff}}, \bibinfo {author}
  {\bibfnamefont {M.}~\bibnamefont {Brune}}, \bibinfo {author} {\bibfnamefont
  {J.-M.}\ \bibnamefont {Raimond}}, \ and\ \bibinfo {author} {\bibfnamefont
  {S.}~\bibnamefont {Haroche}},\ }\href@noop {} {\bibfield  {journal} {\bibinfo
   {journal} {Nature}\ }\textbf {\bibinfo {volume} {446}},\ \bibinfo {pages}
  {297} (\bibinfo {year} {2007})}\BibitemShut {NoStop}%
\bibitem [{\citenamefont {Sayrin}\ \emph {et~al.}(2011)\citenamefont {Sayrin},
  \citenamefont {Dotsenko}, \citenamefont {Zhou}, \citenamefont {Peaudecerf},
  \citenamefont {Rybarczyk}, \citenamefont {Gleyzes}, \citenamefont {Rouchon},
  \citenamefont {Mirrahimi}, \citenamefont {Amini}, \citenamefont {Brune},
  \citenamefont {Raimond},\ and\ \citenamefont {Haroche}}]{Sayrin2011}%
  \BibitemOpen
  \bibfield  {author} {\bibinfo {author} {\bibfnamefont {C.}~\bibnamefont
  {Sayrin}}, \bibinfo {author} {\bibfnamefont {I.}~\bibnamefont {Dotsenko}},
  \bibinfo {author} {\bibfnamefont {X.}~\bibnamefont {Zhou}}, \bibinfo {author}
  {\bibfnamefont {B.}~\bibnamefont {Peaudecerf}}, \bibinfo {author}
  {\bibfnamefont {T.}~\bibnamefont {Rybarczyk}}, \bibinfo {author}
  {\bibfnamefont {S.}~\bibnamefont {Gleyzes}}, \bibinfo {author} {\bibfnamefont
  {P.}~\bibnamefont {Rouchon}}, \bibinfo {author} {\bibfnamefont
  {M.}~\bibnamefont {Mirrahimi}}, \bibinfo {author} {\bibfnamefont
  {H.}~\bibnamefont {Amini}}, \bibinfo {author} {\bibfnamefont
  {M.}~\bibnamefont {Brune}}, \bibinfo {author} {\bibfnamefont {J.-M.}\
  \bibnamefont {Raimond}}, \ and\ \bibinfo {author} {\bibfnamefont
  {S.}~\bibnamefont {Haroche}},\ }\href@noop {} {\bibfield  {journal} {\bibinfo
   {journal} {Nature}\ }\textbf {\bibinfo {volume} {477}},\ \bibinfo {pages}
  {73} (\bibinfo {year} {2011})}\BibitemShut {NoStop}%
\bibitem [{\citenamefont {Zhou}\ \emph
  {et~al.}(2012{\natexlab{a}})\citenamefont {Zhou}, \citenamefont {Dotsenko},
  \citenamefont {Peaudecerf}, \citenamefont {Rybarczyk}, \citenamefont
  {Sayrin}, \citenamefont {Gleyzes}, \citenamefont {Raimond}, \citenamefont
  {Brune},\ and\ \citenamefont {Haroche}}]{Zhou2012b}%
  \BibitemOpen
  \bibfield  {author} {\bibinfo {author} {\bibfnamefont {X.}~\bibnamefont
  {Zhou}}, \bibinfo {author} {\bibfnamefont {I.}~\bibnamefont {Dotsenko}},
  \bibinfo {author} {\bibfnamefont {B.}~\bibnamefont {Peaudecerf}}, \bibinfo
  {author} {\bibfnamefont {T.}~\bibnamefont {Rybarczyk}}, \bibinfo {author}
  {\bibfnamefont {C.}~\bibnamefont {Sayrin}}, \bibinfo {author} {\bibfnamefont
  {S.}~\bibnamefont {Gleyzes}}, \bibinfo {author} {\bibfnamefont {J.~M.}\
  \bibnamefont {Raimond}}, \bibinfo {author} {\bibfnamefont {M.}~\bibnamefont
  {Brune}}, \ and\ \bibinfo {author} {\bibfnamefont {S.}~\bibnamefont
  {Haroche}},\ }\href@noop {} {\bibfield  {journal} {\bibinfo  {journal} {Phys.
  Rev. Lett.}\ }\textbf {\bibinfo {volume} {108}},\ \bibinfo {pages} {243602}
  (\bibinfo {year} {2012}{\natexlab{a}})}\BibitemShut {NoStop}%
\bibitem [{\citenamefont {Kwiat}\ \emph {et~al.}(2001)\citenamefont {Kwiat},
  \citenamefont {Barraza-Lopez}, \citenamefont {Stefanov},\ and\ \citenamefont
  {Gisin}}]{Kwiat2001}%
  \BibitemOpen
  \bibfield  {author} {\bibinfo {author} {\bibfnamefont {P.~G.}\ \bibnamefont
  {Kwiat}}, \bibinfo {author} {\bibfnamefont {S.}~\bibnamefont
  {Barraza-Lopez}}, \bibinfo {author} {\bibfnamefont {A.}~\bibnamefont
  {Stefanov}}, \ and\ \bibinfo {author} {\bibfnamefont {N.}~\bibnamefont
  {Gisin}},\ }\href@noop {} {\bibfield  {journal} {\bibinfo  {journal}
  {Nature}\ }\textbf {\bibinfo {volume} {409}},\ \bibinfo {pages} {1014}
  (\bibinfo {year} {2001})}\BibitemShut {NoStop}%
\bibitem [{\citenamefont {Ritchie}\ \emph {et~al.}(1991)\citenamefont
  {Ritchie}, \citenamefont {Story},\ and\ \citenamefont {Hulet}}]{Ritchie1991}%
  \BibitemOpen
  \bibfield  {author} {\bibinfo {author} {\bibfnamefont {N.~W.~M.}\
  \bibnamefont {Ritchie}}, \bibinfo {author} {\bibfnamefont {J.~G.}\
  \bibnamefont {Story}}, \ and\ \bibinfo {author} {\bibfnamefont {R.~G.}\
  \bibnamefont {Hulet}},\ }\href@noop {} {\bibfield  {journal} {\bibinfo
  {journal} {Phys. Rev. Lett.}\ }\textbf {\bibinfo {volume} {66}},\ \bibinfo
  {pages} {1107} (\bibinfo {year} {1991})}\BibitemShut {NoStop}%
\bibitem [{\citenamefont {Pryde}\ \emph {et~al.}(2005)\citenamefont {Pryde},
  \citenamefont {O'Brien}, \citenamefont {White}, \citenamefont {Ralph},\ and\
  \citenamefont {Wiseman}}]{Pryde2005}%
  \BibitemOpen
  \bibfield  {author} {\bibinfo {author} {\bibfnamefont {G.~J.}\ \bibnamefont
  {Pryde}}, \bibinfo {author} {\bibfnamefont {J.~L.}\ \bibnamefont {O'Brien}},
  \bibinfo {author} {\bibfnamefont {A.~G.}\ \bibnamefont {White}}, \bibinfo
  {author} {\bibfnamefont {T.~C.}\ \bibnamefont {Ralph}}, \ and\ \bibinfo
  {author} {\bibfnamefont {H.~M.}\ \bibnamefont {Wiseman}},\ }\href@noop {}
  {\bibfield  {journal} {\bibinfo  {journal} {Phys. Rev. Lett.}\ }\textbf
  {\bibinfo {volume} {94}},\ \bibinfo {pages} {220405} (\bibinfo {year}
  {2005})}\BibitemShut {NoStop}%
\bibitem [{\citenamefont {Iinuma}\ \emph {et~al.}(2011)\citenamefont {Iinuma},
  \citenamefont {Suzuki}, \citenamefont {Taguchi}, \citenamefont {Kadoya},\
  and\ \citenamefont {Hofmann}}]{Iinuma2011}%
  \BibitemOpen
  \bibfield  {author} {\bibinfo {author} {\bibfnamefont {M.}~\bibnamefont
  {Iinuma}}, \bibinfo {author} {\bibfnamefont {Y.}~\bibnamefont {Suzuki}},
  \bibinfo {author} {\bibfnamefont {G.}~\bibnamefont {Taguchi}}, \bibinfo
  {author} {\bibfnamefont {Y.}~\bibnamefont {Kadoya}}, \ and\ \bibinfo {author}
  {\bibfnamefont {H.~F.}\ \bibnamefont {Hofmann}},\ }\href@noop {} {\bibfield
  {journal} {\bibinfo  {journal} {New J. Phys.}\ }\textbf {\bibinfo {volume}
  {13}},\ \bibinfo {pages} {033041} (\bibinfo {year} {2011})}\BibitemShut
  {NoStop}%
\bibitem [{\citenamefont {Goggin}\ \emph {et~al.}(2011)\citenamefont {Goggin},
  \citenamefont {Almeida}, \citenamefont {Barbieri}, \citenamefont {Lanyon},
  \citenamefont {O'Brien}, \citenamefont {White},\ and\ \citenamefont
  {Pryde}}]{Goggin2011}%
  \BibitemOpen
  \bibfield  {author} {\bibinfo {author} {\bibfnamefont {M.~E.}\ \bibnamefont
  {Goggin}}, \bibinfo {author} {\bibfnamefont {M.~P.}\ \bibnamefont {Almeida}},
  \bibinfo {author} {\bibfnamefont {M.}~\bibnamefont {Barbieri}}, \bibinfo
  {author} {\bibfnamefont {B.~P.}\ \bibnamefont {Lanyon}}, \bibinfo {author}
  {\bibfnamefont {J.~L.}\ \bibnamefont {O'Brien}}, \bibinfo {author}
  {\bibfnamefont {A.~G.}\ \bibnamefont {White}}, \ and\ \bibinfo {author}
  {\bibfnamefont {G.~J.}\ \bibnamefont {Pryde}},\ }\href@noop {} {\bibfield
  {journal} {\bibinfo  {journal} {Proc. Natl. Acad. Sci. U. S. A.}\ }\textbf
  {\bibinfo {volume} {108}},\ \bibinfo {pages} {1256} (\bibinfo {year}
  {2011})}\BibitemShut {NoStop}%
\bibitem [{\citenamefont {Dressel}\ \emph {et~al.}(2011)\citenamefont
  {Dressel}, \citenamefont {Broadbent}, \citenamefont {Howell},\ and\
  \citenamefont {Jordan}}]{Dressel2011}%
  \BibitemOpen
  \bibfield  {author} {\bibinfo {author} {\bibfnamefont {J.}~\bibnamefont
  {Dressel}}, \bibinfo {author} {\bibfnamefont {C.~J.}\ \bibnamefont
  {Broadbent}}, \bibinfo {author} {\bibfnamefont {J.~C.}\ \bibnamefont
  {Howell}}, \ and\ \bibinfo {author} {\bibfnamefont {A.~N.}\ \bibnamefont
  {Jordan}},\ }\href@noop {} {\bibfield  {journal} {\bibinfo  {journal} {Phys.
  Rev. Lett.}\ }\textbf {\bibinfo {volume} {106}},\ \bibinfo {pages} {040402}
  (\bibinfo {year} {2011})}\BibitemShut {NoStop}%
\bibitem [{\citenamefont {Suzuki}\ \emph {et~al.}(2012)\citenamefont {Suzuki},
  \citenamefont {Iinuma},\ and\ \citenamefont {Hofmann}}]{Suzuki2012}%
  \BibitemOpen
  \bibfield  {author} {\bibinfo {author} {\bibfnamefont {Y.}~\bibnamefont
  {Suzuki}}, \bibinfo {author} {\bibfnamefont {M.}~\bibnamefont {Iinuma}}, \
  and\ \bibinfo {author} {\bibfnamefont {H.~F.}\ \bibnamefont {Hofmann}},\
  }\href@noop {} {\bibfield  {journal} {\bibinfo  {journal} {New J. Phys.}\
  }\textbf {\bibinfo {volume} {14}},\ \bibinfo {pages} {103022} (\bibinfo
  {year} {2012})}\BibitemShut {NoStop}%
\bibitem [{\citenamefont {Hosten}\ and\ \citenamefont
  {Kwiat}(2008)}]{Hosten2008}%
  \BibitemOpen
  \bibfield  {author} {\bibinfo {author} {\bibfnamefont {O.}~\bibnamefont
  {Hosten}}\ and\ \bibinfo {author} {\bibfnamefont {P.}~\bibnamefont {Kwiat}},\
  }\href@noop {} {\bibfield  {journal} {\bibinfo  {journal} {Science}\ }\textbf
  {\bibinfo {volume} {319}},\ \bibinfo {pages} {787} (\bibinfo {year}
  {2008})}\BibitemShut {NoStop}%
\bibitem [{\citenamefont {Dixon}\ \emph {et~al.}(2009)\citenamefont {Dixon},
  \citenamefont {Starling}, \citenamefont {Jordan},\ and\ \citenamefont
  {Howell}}]{Dixon2009}%
  \BibitemOpen
  \bibfield  {author} {\bibinfo {author} {\bibfnamefont {P.~B.}\ \bibnamefont
  {Dixon}}, \bibinfo {author} {\bibfnamefont {D.~J.}\ \bibnamefont {Starling}},
  \bibinfo {author} {\bibfnamefont {A.~N.}\ \bibnamefont {Jordan}}, \ and\
  \bibinfo {author} {\bibfnamefont {J.~C.}\ \bibnamefont {Howell}},\
  }\href@noop {} {\bibfield  {journal} {\bibinfo  {journal} {Phys. Rev. Lett.}\
  }\textbf {\bibinfo {volume} {102}},\ \bibinfo {pages} {173601} (\bibinfo
  {year} {2009})}\BibitemShut {NoStop}%
\bibitem [{\citenamefont {Starling}\ \emph {et~al.}(2010)\citenamefont
  {Starling}, \citenamefont {Dixon}, \citenamefont {Williams}, \citenamefont
  {Jordan},\ and\ \citenamefont {Howell}}]{Starling2010b}%
  \BibitemOpen
  \bibfield  {author} {\bibinfo {author} {\bibfnamefont {D.~J.}\ \bibnamefont
  {Starling}}, \bibinfo {author} {\bibfnamefont {P.~B.}\ \bibnamefont {Dixon}},
  \bibinfo {author} {\bibfnamefont {N.~S.}\ \bibnamefont {Williams}}, \bibinfo
  {author} {\bibfnamefont {A.~N.}\ \bibnamefont {Jordan}}, \ and\ \bibinfo
  {author} {\bibfnamefont {J.~C.}\ \bibnamefont {Howell}},\ }\href@noop {}
  {\bibfield  {journal} {\bibinfo  {journal} {Phys. Rev. A}\ }\textbf {\bibinfo
  {volume} {82}},\ \bibinfo {pages} {011802(R)} (\bibinfo {year}
  {2010})}\BibitemShut {NoStop}%
\bibitem [{\citenamefont {Turner}\ \emph {et~al.}(2011)\citenamefont {Turner},
  \citenamefont {Hagedorn}, \citenamefont {Schlamminger},\ and\ \citenamefont
  {Gundlach}}]{Turner2011}%
  \BibitemOpen
  \bibfield  {author} {\bibinfo {author} {\bibfnamefont {M.~D.}\ \bibnamefont
  {Turner}}, \bibinfo {author} {\bibfnamefont {C.~A.}\ \bibnamefont
  {Hagedorn}}, \bibinfo {author} {\bibfnamefont {S.}~\bibnamefont
  {Schlamminger}}, \ and\ \bibinfo {author} {\bibfnamefont {J.~H.}\
  \bibnamefont {Gundlach}},\ }\href@noop {} {\bibfield  {journal} {\bibinfo
  {journal} {Opt. Lett.}\ }\textbf {\bibinfo {volume} {36}},\ \bibinfo {pages}
  {1479} (\bibinfo {year} {2011})}\BibitemShut {NoStop}%
\bibitem [{\citenamefont {Zhou}\ \emph
  {et~al.}(2012{\natexlab{b}})\citenamefont {Zhou}, \citenamefont {Xiao},
  \citenamefont {Luo},\ and\ \citenamefont {Wen}}]{Zhou2012}%
  \BibitemOpen
  \bibfield  {author} {\bibinfo {author} {\bibfnamefont {X.}~\bibnamefont
  {Zhou}}, \bibinfo {author} {\bibfnamefont {Z.}~\bibnamefont {Xiao}}, \bibinfo
  {author} {\bibfnamefont {H.}~\bibnamefont {Luo}}, \ and\ \bibinfo {author}
  {\bibfnamefont {S.}~\bibnamefont {Wen}},\ }\href@noop {} {\bibfield
  {journal} {\bibinfo  {journal} {Phys. Rev. A}\ }\textbf {\bibinfo {volume}
  {85}},\ \bibinfo {pages} {043809} (\bibinfo {year}
  {2012}{\natexlab{b}})}\BibitemShut {NoStop}%
\bibitem [{\citenamefont {Gorodetski}\ \emph {et~al.}(2012)\citenamefont
  {Gorodetski}, \citenamefont {Bliokh}, \citenamefont {Stein}, \citenamefont
  {Genet}, \citenamefont {Shitrit}, \citenamefont {Kleiner}, \citenamefont
  {Hasman},\ and\ \citenamefont {Ebbesen}}]{Gorodetski2012}%
  \BibitemOpen
  \bibfield  {author} {\bibinfo {author} {\bibfnamefont {Y.}~\bibnamefont
  {Gorodetski}}, \bibinfo {author} {\bibfnamefont {K.~Y.}\ \bibnamefont
  {Bliokh}}, \bibinfo {author} {\bibfnamefont {B.}~\bibnamefont {Stein}},
  \bibinfo {author} {\bibfnamefont {C.}~\bibnamefont {Genet}}, \bibinfo
  {author} {\bibfnamefont {N.}~\bibnamefont {Shitrit}}, \bibinfo {author}
  {\bibfnamefont {V.}~\bibnamefont {Kleiner}}, \bibinfo {author} {\bibfnamefont
  {E.}~\bibnamefont {Hasman}}, \ and\ \bibinfo {author} {\bibfnamefont {T.~W.}\
  \bibnamefont {Ebbesen}},\ }\href@noop {} {\bibfield  {journal} {\bibinfo
  {journal} {Phys. Rev. Lett.}\ }\textbf {\bibinfo {volume} {109}},\ \bibinfo
  {pages} {013901} (\bibinfo {year} {2012})}\BibitemShut {NoStop}%
\bibitem [{\citenamefont {Zhou}\ \emph {et~al.}()\citenamefont {Zhou},
  \citenamefont {Turek}, \citenamefont {Sun},\ and\ \citenamefont
  {Nori}}]{Zhou2013}%
  \BibitemOpen
  \bibfield  {author} {\bibinfo {author} {\bibfnamefont {L.}~\bibnamefont
  {Zhou}}, \bibinfo {author} {\bibfnamefont {Y.}~\bibnamefont {Turek}},
  \bibinfo {author} {\bibfnamefont {C.~P.}\ \bibnamefont {Sun}}, \ and\
  \bibinfo {author} {\bibfnamefont {F.}~\bibnamefont {Nori}},\ }\href@noop {}
  {}\Eprint {http://arxiv.org/abs/arXiv:1302.0455} {arXiv:1302.0455}
  \BibitemShut {NoStop}%
\bibitem [{\citenamefont {Kocsis}\ \emph {et~al.}(2011)\citenamefont {Kocsis},
  \citenamefont {Braverman}, \citenamefont {Ravets}, \citenamefont {Stevens},
  \citenamefont {Mirin}, \citenamefont {Shalm},\ and\ \citenamefont
  {Steinberg}}]{Kocsis2011}%
  \BibitemOpen
  \bibfield  {author} {\bibinfo {author} {\bibfnamefont {S.}~\bibnamefont
  {Kocsis}}, \bibinfo {author} {\bibfnamefont {B.}~\bibnamefont {Braverman}},
  \bibinfo {author} {\bibfnamefont {S.}~\bibnamefont {Ravets}}, \bibinfo
  {author} {\bibfnamefont {M.~J.}\ \bibnamefont {Stevens}}, \bibinfo {author}
  {\bibfnamefont {R.~P.}\ \bibnamefont {Mirin}}, \bibinfo {author}
  {\bibfnamefont {L.~K.}\ \bibnamefont {Shalm}}, \ and\ \bibinfo {author}
  {\bibfnamefont {A.~M.}\ \bibnamefont {Steinberg}},\ }\href@noop {} {\bibfield
   {journal} {\bibinfo  {journal} {Science}\ }\textbf {\bibinfo {volume}
  {332}},\ \bibinfo {pages} {1170} (\bibinfo {year} {2011})}\BibitemShut
  {NoStop}%
\bibitem [{\citenamefont {Lundeen}\ \emph {et~al.}(2011)\citenamefont
  {Lundeen}, \citenamefont {Sutherland}, \citenamefont {Patel}, \citenamefont
  {Stewart},\ and\ \citenamefont {Bamber}}]{Lundeen2011}%
  \BibitemOpen
  \bibfield  {author} {\bibinfo {author} {\bibfnamefont {J.~S.}\ \bibnamefont
  {Lundeen}}, \bibinfo {author} {\bibfnamefont {B.}~\bibnamefont {Sutherland}},
  \bibinfo {author} {\bibfnamefont {A.}~\bibnamefont {Patel}}, \bibinfo
  {author} {\bibfnamefont {C.}~\bibnamefont {Stewart}}, \ and\ \bibinfo
  {author} {\bibfnamefont {C.}~\bibnamefont {Bamber}},\ }\href@noop {}
  {\bibfield  {journal} {\bibinfo  {journal} {Nature}\ }\textbf {\bibinfo
  {volume} {474}},\ \bibinfo {pages} {188} (\bibinfo {year}
  {2011})}\BibitemShut {NoStop}%
\bibitem [{\citenamefont {Rozema}\ \emph {et~al.}(2012)\citenamefont {Rozema},
  \citenamefont {Darabi}, \citenamefont {Mahler}, \citenamefont {Hayat},
  \citenamefont {Soudagar},\ and\ \citenamefont {Steinberg}}]{Rozema2012}%
  \BibitemOpen
  \bibfield  {author} {\bibinfo {author} {\bibfnamefont {L.~A.}\ \bibnamefont
  {Rozema}}, \bibinfo {author} {\bibfnamefont {A.}~\bibnamefont {Darabi}},
  \bibinfo {author} {\bibfnamefont {D.~H.}\ \bibnamefont {Mahler}}, \bibinfo
  {author} {\bibfnamefont {A.}~\bibnamefont {Hayat}}, \bibinfo {author}
  {\bibfnamefont {Y.}~\bibnamefont {Soudagar}}, \ and\ \bibinfo {author}
  {\bibfnamefont {A.~M.}\ \bibnamefont {Steinberg}},\ }\href@noop {} {\bibfield
   {journal} {\bibinfo  {journal} {Phys. Rev. Lett.}\ }\textbf {\bibinfo
  {volume} {109}},\ \bibinfo {pages} {100404} (\bibinfo {year}
  {2012})}\BibitemShut {NoStop}%
\bibitem [{\citenamefont {Kaneda}\ \emph {et~al.}()\citenamefont {Kaneda},
  \citenamefont {Baek}, \citenamefont {Ozawa},\ and\ \citenamefont
  {Edamatsu}}]{Kaneda2013}%
  \BibitemOpen
  \bibfield  {author} {\bibinfo {author} {\bibfnamefont {F.}~\bibnamefont
  {Kaneda}}, \bibinfo {author} {\bibfnamefont {S.-Y.}\ \bibnamefont {Baek}},
  \bibinfo {author} {\bibfnamefont {M.}~\bibnamefont {Ozawa}}, \ and\ \bibinfo
  {author} {\bibfnamefont {K.}~\bibnamefont {Edamatsu}},\ }\href@noop {}
  {}\Eprint {http://arxiv.org/abs/arXiv:1308.5868} {arXiv:1308.5868}
  \BibitemShut {NoStop}%
\bibitem [{\citenamefont {Kim}\ \emph {et~al.}(2009)\citenamefont {Kim},
  \citenamefont {Cho}, \citenamefont {Ra},\ and\ \citenamefont
  {Kim}}]{Kim2009}%
  \BibitemOpen
  \bibfield  {author} {\bibinfo {author} {\bibfnamefont {Y.-S.}\ \bibnamefont
  {Kim}}, \bibinfo {author} {\bibfnamefont {Y.-W.}\ \bibnamefont {Cho}},
  \bibinfo {author} {\bibfnamefont {Y.-S.}\ \bibnamefont {Ra}}, \ and\ \bibinfo
  {author} {\bibfnamefont {Y.-H.}\ \bibnamefont {Kim}},\ }\href@noop {}
  {\bibfield  {journal} {\bibinfo  {journal} {Opt. Exp.}\ }\textbf {\bibinfo
  {volume} {17}},\ \bibinfo {pages} {11978} (\bibinfo {year}
  {2009})}\BibitemShut {NoStop}%
\bibitem [{\citenamefont {Kim}\ \emph {et~al.}(2012)\citenamefont {Kim},
  \citenamefont {Lee}, \citenamefont {Kwon},\ and\ \citenamefont
  {Kim}}]{Kim2012}%
  \BibitemOpen
  \bibfield  {author} {\bibinfo {author} {\bibfnamefont {Y.-S.}\ \bibnamefont
  {Kim}}, \bibinfo {author} {\bibfnamefont {J.-C.}\ \bibnamefont {Lee}},
  \bibinfo {author} {\bibfnamefont {O.}~\bibnamefont {Kwon}}, \ and\ \bibinfo
  {author} {\bibfnamefont {Y.-H.}\ \bibnamefont {Kim}},\ }\href@noop {}
  {\bibfield  {journal} {\bibinfo  {journal} {Nature Phys.}\ }\textbf {\bibinfo
  {volume} {8}},\ \bibinfo {pages} {117} (\bibinfo {year} {2012})}\BibitemShut
  {NoStop}%
\bibitem [{\citenamefont {Lundeen}\ and\ \citenamefont
  {Steinberg}(2009)}]{Lundeen2009}%
  \BibitemOpen
  \bibfield  {author} {\bibinfo {author} {\bibfnamefont {J.~S.}\ \bibnamefont
  {Lundeen}}\ and\ \bibinfo {author} {\bibfnamefont {A.~M.}\ \bibnamefont
  {Steinberg}},\ }\href@noop {} {\bibfield  {journal} {\bibinfo  {journal}
  {Phys. Rev. Lett.}\ }\textbf {\bibinfo {volume} {102}},\ \bibinfo {pages}
  {020404} (\bibinfo {year} {2009})}\BibitemShut {NoStop}%
\bibitem [{\citenamefont {Yokota}\ \emph {et~al.}(2009)\citenamefont {Yokota},
  \citenamefont {Yamamoto}, \citenamefont {Koashi},\ and\ \citenamefont
  {Imoto}}]{Yokota2009}%
  \BibitemOpen
  \bibfield  {author} {\bibinfo {author} {\bibfnamefont {K.}~\bibnamefont
  {Yokota}}, \bibinfo {author} {\bibfnamefont {T.}~\bibnamefont {Yamamoto}},
  \bibinfo {author} {\bibfnamefont {M.}~\bibnamefont {Koashi}}, \ and\ \bibinfo
  {author} {\bibfnamefont {N.}~\bibnamefont {Imoto}},\ }\href@noop {}
  {\bibfield  {journal} {\bibinfo  {journal} {New J. Phys.}\ }\textbf {\bibinfo
  {volume} {11}},\ \bibinfo {pages} {033011} (\bibinfo {year}
  {2009})}\BibitemShut {NoStop}%
\bibitem [{\citenamefont {James}\ \emph {et~al.}(2001)\citenamefont {James},
  \citenamefont {Kwiat}, \citenamefont {Munro},\ and\ \citenamefont
  {White}}]{James2001}%
  \BibitemOpen
  \bibfield  {author} {\bibinfo {author} {\bibfnamefont {D.~F.~V.}\
  \bibnamefont {James}}, \bibinfo {author} {\bibfnamefont {P.~G.}\ \bibnamefont
  {Kwiat}}, \bibinfo {author} {\bibfnamefont {W.~J.}\ \bibnamefont {Munro}}, \
  and\ \bibinfo {author} {\bibfnamefont {A.~G.}\ \bibnamefont {White}},\
  }\href@noop {} {\bibfield  {journal} {\bibinfo  {journal} {Phys. Rev. A}\
  }\textbf {\bibinfo {volume} {64}},\ \bibinfo {pages} {052312} (\bibinfo
  {year} {2001})}\BibitemShut {NoStop}%
\bibitem [{\citenamefont {Andersson}\ and\ \citenamefont
  {Oi}(2008)}]{Andersson2008}%
  \BibitemOpen
  \bibfield  {author} {\bibinfo {author} {\bibfnamefont {E.}~\bibnamefont
  {Andersson}}\ and\ \bibinfo {author} {\bibfnamefont {D.~K.~L.}\ \bibnamefont
  {Oi}},\ }\href@noop {} {\bibfield  {journal} {\bibinfo  {journal} {Phys. Rev.
  A}\ }\textbf {\bibinfo {volume} {77}},\ \bibinfo {pages} {052104} (\bibinfo
  {year} {2008})}\BibitemShut {NoStop}%
\bibitem [{\citenamefont {Ota}\ \emph {et~al.}(2012)\citenamefont {Ota},
  \citenamefont {Ashhab},\ and\ \citenamefont {Nori}}]{Ota2012}%
  \BibitemOpen
  \bibfield  {author} {\bibinfo {author} {\bibfnamefont {Y.}~\bibnamefont
  {Ota}}, \bibinfo {author} {\bibfnamefont {S.}~\bibnamefont {Ashhab}}, \ and\
  \bibinfo {author} {\bibfnamefont {F.}~\bibnamefont {Nori}},\ }\href@noop {}
  {\bibfield  {journal} {\bibinfo  {journal} {Phys. Rev. A}\ }\textbf {\bibinfo
  {volume} {85}},\ \bibinfo {pages} {043808} (\bibinfo {year}
  {2012})}\BibitemShut {NoStop}%
\bibitem [{\citenamefont {Katz}\ \emph {et~al.}(2006)\citenamefont {Katz},
  \citenamefont {Ansmann}, \citenamefont {Bialczak}, \citenamefont {Lucero},
  \citenamefont {McDermott}, \citenamefont {Neeley}, \citenamefont {Steffen},
  \citenamefont {Weig}, \citenamefont {Cleland}, \citenamefont {Martinis},\
  and\ \citenamefont {Korotkov}}]{Katz2006}%
  \BibitemOpen
  \bibfield  {author} {\bibinfo {author} {\bibfnamefont {N.}~\bibnamefont
  {Katz}}, \bibinfo {author} {\bibfnamefont {M.}~\bibnamefont {Ansmann}},
  \bibinfo {author} {\bibfnamefont {R.~C.}\ \bibnamefont {Bialczak}}, \bibinfo
  {author} {\bibfnamefont {E.}~\bibnamefont {Lucero}}, \bibinfo {author}
  {\bibfnamefont {R.}~\bibnamefont {McDermott}}, \bibinfo {author}
  {\bibfnamefont {M.}~\bibnamefont {Neeley}}, \bibinfo {author} {\bibfnamefont
  {M.}~\bibnamefont {Steffen}}, \bibinfo {author} {\bibfnamefont {E.~M.}\
  \bibnamefont {Weig}}, \bibinfo {author} {\bibfnamefont {A.~N.}\ \bibnamefont
  {Cleland}}, \bibinfo {author} {\bibfnamefont {J.~M.}\ \bibnamefont
  {Martinis}}, \ and\ \bibinfo {author} {\bibfnamefont {A.~N.}\ \bibnamefont
  {Korotkov}},\ }\href@noop {} {\bibfield  {journal} {\bibinfo  {journal}
  {Science}\ }\textbf {\bibinfo {volume} {312}},\ \bibinfo {pages} {1498}
  (\bibinfo {year} {2006})}\BibitemShut {NoStop}%
\bibitem [{\citenamefont {Katz}\ \emph {et~al.}(2008)\citenamefont {Katz},
  \citenamefont {Neeley}, \citenamefont {Ansmann}, \citenamefont {Bialczak},
  \citenamefont {Hofheinz}, \citenamefont {Lucero}, \citenamefont {O'Connell},
  \citenamefont {Wang}, \citenamefont {Cleland}, \citenamefont {Martinis},\
  and\ \citenamefont {Korotkov}}]{Katz2008}%
  \BibitemOpen
  \bibfield  {author} {\bibinfo {author} {\bibfnamefont {N.}~\bibnamefont
  {Katz}}, \bibinfo {author} {\bibfnamefont {M.}~\bibnamefont {Neeley}},
  \bibinfo {author} {\bibfnamefont {M.}~\bibnamefont {Ansmann}}, \bibinfo
  {author} {\bibfnamefont {R.}~\bibnamefont {Bialczak}}, \bibinfo {author}
  {\bibfnamefont {M.}~\bibnamefont {Hofheinz}}, \bibinfo {author}
  {\bibfnamefont {E.}~\bibnamefont {Lucero}}, \bibinfo {author} {\bibfnamefont
  {A.}~\bibnamefont {O'Connell}}, \bibinfo {author} {\bibfnamefont
  {H.}~\bibnamefont {Wang}}, \bibinfo {author} {\bibfnamefont {A.}~\bibnamefont
  {Cleland}}, \bibinfo {author} {\bibfnamefont {J.}~\bibnamefont {Martinis}}, \
  and\ \bibinfo {author} {\bibfnamefont {A.}~\bibnamefont {Korotkov}},\
  }\href@noop {} {\bibfield  {journal} {\bibinfo  {journal} {Phys. Rev. Lett.}\
  }\textbf {\bibinfo {volume} {101}},\ \bibinfo {pages} {200401} (\bibinfo
  {year} {2008})}\BibitemShut {NoStop}%
\bibitem [{\citenamefont {Palacios-Laloy}\ \emph {et~al.}(2010)\citenamefont
  {Palacios-Laloy}, \citenamefont {Mallet}, \citenamefont {Nguyen},
  \citenamefont {Bertet}, \citenamefont {Vion}, \citenamefont {Esteve},\ and\
  \citenamefont {Korotkov}}]{Palacios-Laloy2010}%
  \BibitemOpen
  \bibfield  {author} {\bibinfo {author} {\bibfnamefont {A.}~\bibnamefont
  {Palacios-Laloy}}, \bibinfo {author} {\bibfnamefont {F.}~\bibnamefont
  {Mallet}}, \bibinfo {author} {\bibfnamefont {F.}~\bibnamefont {Nguyen}},
  \bibinfo {author} {\bibfnamefont {P.}~\bibnamefont {Bertet}}, \bibinfo
  {author} {\bibfnamefont {D.}~\bibnamefont {Vion}}, \bibinfo {author}
  {\bibfnamefont {D.}~\bibnamefont {Esteve}}, \ and\ \bibinfo {author}
  {\bibfnamefont {A.~N.}\ \bibnamefont {Korotkov}},\ }\href@noop {} {\bibfield
  {journal} {\bibinfo  {journal} {Nature Phys.}\ }\textbf {\bibinfo {volume}
  {6}},\ \bibinfo {pages} {442} (\bibinfo {year} {2010})}\BibitemShut {NoStop}%
\bibitem [{\citenamefont {Groen}\ \emph {et~al.}(2013)\citenamefont {Groen},
  \citenamefont {Rist\`{e}}, \citenamefont {Tornberg}, \citenamefont {Cramer},
  \citenamefont {de~Groot}, \citenamefont {Picot}, \citenamefont {Johansson},\
  and\ \citenamefont {DiCarlo}}]{Groen2013}%
  \BibitemOpen
  \bibfield  {author} {\bibinfo {author} {\bibfnamefont {J.~P.}\ \bibnamefont
  {Groen}}, \bibinfo {author} {\bibfnamefont {D.}~\bibnamefont {Rist\`{e}}},
  \bibinfo {author} {\bibfnamefont {L.}~\bibnamefont {Tornberg}}, \bibinfo
  {author} {\bibfnamefont {J.}~\bibnamefont {Cramer}}, \bibinfo {author}
  {\bibfnamefont {P.~C.}\ \bibnamefont {de~Groot}}, \bibinfo {author}
  {\bibfnamefont {T.}~\bibnamefont {Picot}}, \bibinfo {author} {\bibfnamefont
  {G.}~\bibnamefont {Johansson}}, \ and\ \bibinfo {author} {\bibfnamefont
  {L.}~\bibnamefont {DiCarlo}},\ }\href@noop {} {\bibfield  {journal} {\bibinfo
   {journal} {Phys. Rev. Lett.}\ }\textbf {\bibinfo {volume} {111}},\ \bibinfo
  {pages} {090506} (\bibinfo {year} {2013})}\BibitemShut {NoStop}%
\bibitem [{\citenamefont {Vijay}\ \emph {et~al.}(2012)\citenamefont {Vijay},
  \citenamefont {Macklin}, \citenamefont {Slichter}, \citenamefont {Weber},
  \citenamefont {Murch}, \citenamefont {Naik}, \citenamefont {Korotkov},\ and\
  \citenamefont {Siddiqi}}]{Vijay2012}%
  \BibitemOpen
  \bibfield  {author} {\bibinfo {author} {\bibfnamefont {R.}~\bibnamefont
  {Vijay}}, \bibinfo {author} {\bibfnamefont {C.}~\bibnamefont {Macklin}},
  \bibinfo {author} {\bibfnamefont {D.~H.}\ \bibnamefont {Slichter}}, \bibinfo
  {author} {\bibfnamefont {S.~J.}\ \bibnamefont {Weber}}, \bibinfo {author}
  {\bibfnamefont {K.~W.}\ \bibnamefont {Murch}}, \bibinfo {author}
  {\bibfnamefont {R.}~\bibnamefont {Naik}}, \bibinfo {author} {\bibfnamefont
  {A.~N.}\ \bibnamefont {Korotkov}}, \ and\ \bibinfo {author} {\bibfnamefont
  {I.}~\bibnamefont {Siddiqi}},\ }\href@noop {} {\bibfield  {journal} {\bibinfo
   {journal} {Nature}\ }\textbf {\bibinfo {volume} {490}},\ \bibinfo {pages}
  {77} (\bibinfo {year} {2012})}\BibitemShut {NoStop}%
\bibitem [{\citenamefont {Hatridge}\ \emph {et~al.}(2013)\citenamefont
  {Hatridge}, \citenamefont {Shankar}, \citenamefont {Mirrahimi}, \citenamefont
  {Schackert}, \citenamefont {Geerlings}, \citenamefont {Brecht}, \citenamefont
  {Sliwa}, \citenamefont {Abdo}, \citenamefont {Frunzio}, \citenamefont
  {Girvin}, \citenamefont {Schoelkopf},\ and\ \citenamefont
  {Devoret}}]{Hatridge2013}%
  \BibitemOpen
  \bibfield  {author} {\bibinfo {author} {\bibfnamefont {M.~S.}\ \bibnamefont
  {Hatridge}}, \bibinfo {author} {\bibfnamefont {S.}~\bibnamefont {Shankar}},
  \bibinfo {author} {\bibfnamefont {M.}~\bibnamefont {Mirrahimi}}, \bibinfo
  {author} {\bibfnamefont {F.}~\bibnamefont {Schackert}}, \bibinfo {author}
  {\bibfnamefont {K.}~\bibnamefont {Geerlings}}, \bibinfo {author}
  {\bibfnamefont {T.}~\bibnamefont {Brecht}}, \bibinfo {author} {\bibfnamefont
  {K.~M.}\ \bibnamefont {Sliwa}}, \bibinfo {author} {\bibfnamefont
  {B.}~\bibnamefont {Abdo}}, \bibinfo {author} {\bibfnamefont {L.}~\bibnamefont
  {Frunzio}}, \bibinfo {author} {\bibfnamefont {S.~M.}\ \bibnamefont {Girvin}},
  \bibinfo {author} {\bibfnamefont {R.~J.}\ \bibnamefont {Schoelkopf}}, \ and\
  \bibinfo {author} {\bibfnamefont {M.~H.}\ \bibnamefont {Devoret}},\
  }\href@noop {} {\bibfield  {journal} {\bibinfo  {journal} {Science}\ }\textbf
  {\bibinfo {volume} {339}},\ \bibinfo {pages} {178} (\bibinfo {year}
  {2013})}\BibitemShut {NoStop}%
\bibitem [{\citenamefont {Murch}\ \emph {et~al.}(2013)\citenamefont {Murch},
  \citenamefont {Weber}, \citenamefont {Macklin},\ and\ \citenamefont
  {Siddiqi}}]{Murch2013}%
  \BibitemOpen
  \bibfield  {author} {\bibinfo {author} {\bibfnamefont {K.~W.}\ \bibnamefont
  {Murch}}, \bibinfo {author} {\bibfnamefont {S.~J.}\ \bibnamefont {Weber}},
  \bibinfo {author} {\bibfnamefont {C.}~\bibnamefont {Macklin}}, \ and\
  \bibinfo {author} {\bibfnamefont {I.}~\bibnamefont {Siddiqi}},\ }\href@noop
  {} {\bibfield  {journal} {\bibinfo  {journal} {Nature}\ }\textbf {\bibinfo
  {volume} {511}},\ \bibinfo {pages} {211} (\bibinfo {year}
  {2013})}\BibitemShut {NoStop}%
\bibitem [{\citenamefont {Weber}\ \emph {et~al.}()\citenamefont {Weber},
  \citenamefont {Chantasri}, \citenamefont {Dressel}, \citenamefont {Jordan},
  \citenamefont {Murch},\ and\ \citenamefont {Siddiqi}}]{Weber2014}%
  \BibitemOpen
  \bibfield  {author} {\bibinfo {author} {\bibfnamefont {S.~J.}\ \bibnamefont
  {Weber}}, \bibinfo {author} {\bibfnamefont {A.}~\bibnamefont {Chantasri}},
  \bibinfo {author} {\bibfnamefont {J.}~\bibnamefont {Dressel}}, \bibinfo
  {author} {\bibfnamefont {A.~N.}\ \bibnamefont {Jordan}}, \bibinfo {author}
  {\bibfnamefont {K.~W.}\ \bibnamefont {Murch}}, \ and\ \bibinfo {author}
  {\bibfnamefont {I.}~\bibnamefont {Siddiqi}},\ }\href@noop {} {}\Eprint
  {http://arxiv.org/abs/arXiv:1403.4992} {arXiv:1403.4992} \BibitemShut
  {NoStop}%
\bibitem [{\citenamefont {Rist\`e}\ \emph {et~al.}(2013)\citenamefont
  {Rist\`e}, \citenamefont {Dukalski}, \citenamefont {Watson}, \citenamefont
  {de~Lange}, \citenamefont {Tiggelman}, \citenamefont {Blanter}, \citenamefont
  {Lehnert}, \citenamefont {Schouten},\ and\ \citenamefont
  {DiCarlo}}]{Riste2013}%
  \BibitemOpen
  \bibfield  {author} {\bibinfo {author} {\bibfnamefont {D.}~\bibnamefont
  {Rist\`e}}, \bibinfo {author} {\bibfnamefont {M.}~\bibnamefont {Dukalski}},
  \bibinfo {author} {\bibfnamefont {C.~A.}\ \bibnamefont {Watson}}, \bibinfo
  {author} {\bibfnamefont {G.}~\bibnamefont {de~Lange}}, \bibinfo {author}
  {\bibfnamefont {M.~J.}\ \bibnamefont {Tiggelman}}, \bibinfo {author}
  {\bibfnamefont {Y.~M.}\ \bibnamefont {Blanter}}, \bibinfo {author}
  {\bibfnamefont {K.~W.}\ \bibnamefont {Lehnert}}, \bibinfo {author}
  {\bibfnamefont {R.~N.}\ \bibnamefont {Schouten}}, \ and\ \bibinfo {author}
  {\bibfnamefont {L.}~\bibnamefont {DiCarlo}},\ }\href@noop {} {\bibfield
  {journal} {\bibinfo  {journal} {Nature}\ }\textbf {\bibinfo {volume} {502}},\
  \bibinfo {pages} {350} (\bibinfo {year} {2013})}\BibitemShut {NoStop}%
\bibitem [{\citenamefont {Roch}\ \emph {et~al.}(2014)\citenamefont {Roch},
  \citenamefont {Schwartz}, \citenamefont {Motzoi}, \citenamefont {Macklin},
  \citenamefont {Vijay}, \citenamefont {Eddins}, \citenamefont {Korotkov},
  \citenamefont {Whaley}, \citenamefont {Sarovar},\ and\ \citenamefont
  {Siddiqi}}]{Roch-14}%
  \BibitemOpen
  \bibfield  {author} {\bibinfo {author} {\bibfnamefont {N.}~\bibnamefont
  {Roch}}, \bibinfo {author} {\bibfnamefont {M.~E.}\ \bibnamefont {Schwartz}},
  \bibinfo {author} {\bibfnamefont {F.}~\bibnamefont {Motzoi}}, \bibinfo
  {author} {\bibfnamefont {C.}~\bibnamefont {Macklin}}, \bibinfo {author}
  {\bibfnamefont {R.}~\bibnamefont {Vijay}}, \bibinfo {author} {\bibfnamefont
  {A.~W.}\ \bibnamefont {Eddins}}, \bibinfo {author} {\bibfnamefont {A.~N.}\
  \bibnamefont {Korotkov}}, \bibinfo {author} {\bibfnamefont {K.~B.}\
  \bibnamefont {Whaley}}, \bibinfo {author} {\bibfnamefont {M.}~\bibnamefont
  {Sarovar}}, \ and\ \bibinfo {author} {\bibfnamefont {I.}~\bibnamefont
  {Siddiqi}},\ }\href@noop {} {\bibfield  {journal} {\bibinfo  {journal} {Phys.
  Rev. Lett.}\ }\textbf {\bibinfo {volume} {112}},\ \bibinfo {pages} {170501}
  (\bibinfo {year} {2014})}\BibitemShut {NoStop}%
\bibitem [{\citenamefont {Zhong}\ \emph {et~al.}(2014)\citenamefont {Zhong},
  \citenamefont {Wang}, \citenamefont {Martinis}, \citenamefont {Cleland},
  \citenamefont {Korotkov},\ and\ \citenamefont {Wang}}]{Zhong2014}%
  \BibitemOpen
  \bibfield  {author} {\bibinfo {author} {\bibfnamefont {Y.~P.}\ \bibnamefont
  {Zhong}}, \bibinfo {author} {\bibfnamefont {Z.~L.}\ \bibnamefont {Wang}},
  \bibinfo {author} {\bibfnamefont {J.~M.}\ \bibnamefont {Martinis}}, \bibinfo
  {author} {\bibfnamefont {A.~N.}\ \bibnamefont {Cleland}}, \bibinfo {author}
  {\bibfnamefont {A.~N.}\ \bibnamefont {Korotkov}}, \ and\ \bibinfo {author}
  {\bibfnamefont {H.}~\bibnamefont {Wang}},\ }\href@noop {} {\bibfield
  {journal} {\bibinfo  {journal} {Nature Comm.}\ }\textbf {\bibinfo {volume}
  {5}},\ \bibinfo {pages} {3135} (\bibinfo {year} {2014})}\BibitemShut
  {NoStop}%
\bibitem [{\citenamefont {Korotkov}(2008)}]{Korotkov2008}%
  \BibitemOpen
  \bibfield  {author} {\bibinfo {author} {\bibfnamefont {A.~N.}\ \bibnamefont
  {Korotkov}},\ }\href@noop {} {\bibfield  {journal} {\bibinfo  {journal}
  {Phys. Rev. B}\ }\textbf {\bibinfo {volume} {78}},\ \bibinfo {pages} {174512}
  (\bibinfo {year} {2008})}\BibitemShut {NoStop}%
\bibitem [{\citenamefont {Blais}\ \emph {et~al.}(2004)\citenamefont {Blais},
  \citenamefont {Huang}, \citenamefont {Wallraff}, \citenamefont {Girvin},\
  and\ \citenamefont {Schoelkopf}}]{Blais2004}%
  \BibitemOpen
  \bibfield  {author} {\bibinfo {author} {\bibfnamefont {A.}~\bibnamefont
  {Blais}}, \bibinfo {author} {\bibfnamefont {R.~S.}\ \bibnamefont {Huang}},
  \bibinfo {author} {\bibfnamefont {A.}~\bibnamefont {Wallraff}}, \bibinfo
  {author} {\bibfnamefont {S.~M.}\ \bibnamefont {Girvin}}, \ and\ \bibinfo
  {author} {\bibfnamefont {R.~J.}\ \bibnamefont {Schoelkopf}},\ }\href@noop {}
  {\bibfield  {journal} {\bibinfo  {journal} {Phys. Rev. A}\ }\textbf {\bibinfo
  {volume} {69}},\ \bibinfo {pages} {062320} (\bibinfo {year}
  {2004})}\BibitemShut {NoStop}%
\bibitem [{\citenamefont {Wallraff}\ \emph {et~al.}(2004)\citenamefont
  {Wallraff}, \citenamefont {Schuster}, \citenamefont {Blais}, \citenamefont
  {Frunzio}, \citenamefont {Huang}, \citenamefont {Majer}, \citenamefont
  {Kumar}, \citenamefont {Girvin},\ and\ \citenamefont
  {Schoelkopf}}]{Wallraff2004}%
  \BibitemOpen
  \bibfield  {author} {\bibinfo {author} {\bibfnamefont {A.}~\bibnamefont
  {Wallraff}}, \bibinfo {author} {\bibfnamefont {D.~I.}\ \bibnamefont
  {Schuster}}, \bibinfo {author} {\bibfnamefont {A.}~\bibnamefont {Blais}},
  \bibinfo {author} {\bibfnamefont {L.}~\bibnamefont {Frunzio}}, \bibinfo
  {author} {\bibfnamefont {R.~S.}\ \bibnamefont {Huang}}, \bibinfo {author}
  {\bibfnamefont {J.}~\bibnamefont {Majer}}, \bibinfo {author} {\bibfnamefont
  {S.}~\bibnamefont {Kumar}}, \bibinfo {author} {\bibfnamefont {S.~M.}\
  \bibnamefont {Girvin}}, \ and\ \bibinfo {author} {\bibfnamefont {R.~J.}\
  \bibnamefont {Schoelkopf}},\ }\href@noop {} {\bibfield  {journal} {\bibinfo
  {journal} {Nature}\ }\textbf {\bibinfo {volume} {431}},\ \bibinfo {pages}
  {162} (\bibinfo {year} {2004})}\BibitemShut {NoStop}%
\bibitem [{\citenamefont {Chow}\ \emph {et~al.}(2011)\citenamefont {Chow},
  \citenamefont {C\'orcoles}, \citenamefont {Gambetta}, \citenamefont
  {Rigetti}, \citenamefont {Johnson}, \citenamefont {Smolin}, \citenamefont
  {Rozen}, \citenamefont {Keefe}, \citenamefont {Rothwell}, \citenamefont
  {Ketchen},\ and\ \citenamefont {Steffen}}]{Chow2011}%
  \BibitemOpen
  \bibfield  {author} {\bibinfo {author} {\bibfnamefont {J.~M.}\ \bibnamefont
  {Chow}}, \bibinfo {author} {\bibfnamefont {A.~D.}\ \bibnamefont
  {C\'orcoles}}, \bibinfo {author} {\bibfnamefont {J.~M.}\ \bibnamefont
  {Gambetta}}, \bibinfo {author} {\bibfnamefont {C.}~\bibnamefont {Rigetti}},
  \bibinfo {author} {\bibfnamefont {B.~R.}\ \bibnamefont {Johnson}}, \bibinfo
  {author} {\bibfnamefont {J.~A.}\ \bibnamefont {Smolin}}, \bibinfo {author}
  {\bibfnamefont {J.~R.}\ \bibnamefont {Rozen}}, \bibinfo {author}
  {\bibfnamefont {G.~A.}\ \bibnamefont {Keefe}}, \bibinfo {author}
  {\bibfnamefont {M.~B.}\ \bibnamefont {Rothwell}}, \bibinfo {author}
  {\bibfnamefont {M.~B.}\ \bibnamefont {Ketchen}}, \ and\ \bibinfo {author}
  {\bibfnamefont {M.}~\bibnamefont {Steffen}},\ }\href@noop {} {\bibfield
  {journal} {\bibinfo  {journal} {Phys. Rev. Lett.}\ }\textbf {\bibinfo
  {volume} {107}},\ \bibinfo {pages} {080502} (\bibinfo {year}
  {2011})}\BibitemShut {NoStop}%
\bibitem [{\citenamefont {Johnson}\ \emph {et~al.}(2012)\citenamefont
  {Johnson}, \citenamefont {Macklin}, \citenamefont {Slichter}, \citenamefont
  {Vijay}, \citenamefont {Weingarten}, \citenamefont {Clarke},\ and\
  \citenamefont {Siddiqi}}]{Johnson2012}%
  \BibitemOpen
  \bibfield  {author} {\bibinfo {author} {\bibfnamefont {J.~E.}\ \bibnamefont
  {Johnson}}, \bibinfo {author} {\bibfnamefont {C.}~\bibnamefont {Macklin}},
  \bibinfo {author} {\bibfnamefont {D.~H.}\ \bibnamefont {Slichter}}, \bibinfo
  {author} {\bibfnamefont {R.}~\bibnamefont {Vijay}}, \bibinfo {author}
  {\bibfnamefont {E.~B.}\ \bibnamefont {Weingarten}}, \bibinfo {author}
  {\bibfnamefont {J.}~\bibnamefont {Clarke}}, \ and\ \bibinfo {author}
  {\bibfnamefont {I.}~\bibnamefont {Siddiqi}},\ }\href@noop {} {\bibfield
  {journal} {\bibinfo  {journal} {Phys. Rev. Lett.}\ }\textbf {\bibinfo
  {volume} {109}},\ \bibinfo {pages} {050506} (\bibinfo {year}
  {2012})}\BibitemShut {NoStop}%
\bibitem [{\citenamefont {Rist\`e}\ \emph {et~al.}(2012)\citenamefont
  {Rist\`e}, \citenamefont {van Leeuwen}, \citenamefont {Ku}, \citenamefont
  {Lehnert},\ and\ \citenamefont {DiCarlo}}]{Riste2012}%
  \BibitemOpen
  \bibfield  {author} {\bibinfo {author} {\bibfnamefont {D.}~\bibnamefont
  {Rist\`e}}, \bibinfo {author} {\bibfnamefont {J.~G.}\ \bibnamefont {van
  Leeuwen}}, \bibinfo {author} {\bibfnamefont {H.-S.}\ \bibnamefont {Ku}},
  \bibinfo {author} {\bibfnamefont {K.~W.}\ \bibnamefont {Lehnert}}, \ and\
  \bibinfo {author} {\bibfnamefont {L.}~\bibnamefont {DiCarlo}},\ }\href@noop
  {} {\bibfield  {journal} {\bibinfo  {journal} {Phys. Rev. Lett.}\ }\textbf
  {\bibinfo {volume} {109}},\ \bibinfo {pages} {050507} (\bibinfo {year}
  {2012})}\BibitemShut {NoStop}%
\bibitem [{\citenamefont {Barends}\ \emph {et~al.}(2013)\citenamefont
  {Barends}, \citenamefont {Kelly}, \citenamefont {Megrant}, \citenamefont
  {Sank}, \citenamefont {Jeffrey}, \citenamefont {Chen}, \citenamefont {Yin},
  \citenamefont {Chiaro}, \citenamefont {Mutus}, \citenamefont {Neill},
  \citenamefont {O'Malley}, \citenamefont {Roushan}, \citenamefont {Wenner},
  \citenamefont {White}, \citenamefont {Cleland},\ and\ \citenamefont
  {Martinis}}]{Barends2013}%
  \BibitemOpen
  \bibfield  {author} {\bibinfo {author} {\bibfnamefont {R.}~\bibnamefont
  {Barends}}, \bibinfo {author} {\bibfnamefont {J.}~\bibnamefont {Kelly}},
  \bibinfo {author} {\bibfnamefont {A.}~\bibnamefont {Megrant}}, \bibinfo
  {author} {\bibfnamefont {D.}~\bibnamefont {Sank}}, \bibinfo {author}
  {\bibfnamefont {E.}~\bibnamefont {Jeffrey}}, \bibinfo {author} {\bibfnamefont
  {Y.}~\bibnamefont {Chen}}, \bibinfo {author} {\bibfnamefont {Y.}~\bibnamefont
  {Yin}}, \bibinfo {author} {\bibfnamefont {B.}~\bibnamefont {Chiaro}},
  \bibinfo {author} {\bibfnamefont {J.}~\bibnamefont {Mutus}}, \bibinfo
  {author} {\bibfnamefont {C.}~\bibnamefont {Neill}}, \bibinfo {author}
  {\bibfnamefont {P.}~\bibnamefont {O'Malley}}, \bibinfo {author}
  {\bibfnamefont {P.}~\bibnamefont {Roushan}}, \bibinfo {author} {\bibfnamefont
  {J.}~\bibnamefont {Wenner}}, \bibinfo {author} {\bibfnamefont {T.~C.}\
  \bibnamefont {White}}, \bibinfo {author} {\bibfnamefont {A.~N.}\ \bibnamefont
  {Cleland}}, \ and\ \bibinfo {author} {\bibfnamefont {J.~M.}\ \bibnamefont
  {Martinis}},\ }\href@noop {} {\bibfield  {journal} {\bibinfo  {journal}
  {Phys. Rev. Lett.}\ }\textbf {\bibinfo {volume} {111}},\ \bibinfo {pages}
  {080502} (\bibinfo {year} {2013})}\BibitemShut {NoStop}%
\bibitem [{\citenamefont {Gambetta}\ \emph {et~al.}(2008)\citenamefont
  {Gambetta}, \citenamefont {Blais}, \citenamefont {Boissonneault},
  \citenamefont {Houck}, \citenamefont {Schuster},\ and\ \citenamefont
  {Girvin}}]{Gambetta2008}%
  \BibitemOpen
  \bibfield  {author} {\bibinfo {author} {\bibfnamefont {J.}~\bibnamefont
  {Gambetta}}, \bibinfo {author} {\bibfnamefont {A.}~\bibnamefont {Blais}},
  \bibinfo {author} {\bibfnamefont {M.}~\bibnamefont {Boissonneault}}, \bibinfo
  {author} {\bibfnamefont {A.~A.}\ \bibnamefont {Houck}}, \bibinfo {author}
  {\bibfnamefont {D.~I.}\ \bibnamefont {Schuster}}, \ and\ \bibinfo {author}
  {\bibfnamefont {S.~M.}\ \bibnamefont {Girvin}},\ }\href@noop {} {\bibfield
  {journal} {\bibinfo  {journal} {Phys. Rev. A}\ }\textbf {\bibinfo {volume}
  {77}},\ \bibinfo {pages} {012112} (\bibinfo {year} {2008})}\BibitemShut
  {NoStop}%
\bibitem [{\citenamefont {Korotkov}()}]{Korotkov2011}%
  \BibitemOpen
  \bibfield  {author} {\bibinfo {author} {\bibfnamefont {A.~N.}\ \bibnamefont
  {Korotkov}},\ }\href@noop {} {}\bibinfo {note} {{\it Lecture notes of the Les
  Houches Summer School}, Vol. 96, July 2011, Ch. 17 (Oxford Univ. Press,
  2014)},\ \Eprint {http://arxiv.org/abs/arXiv:1111.4016} {arXiv:1111.4016}
  \BibitemShut {NoStop}%
\bibitem [{\citenamefont {Clerk}\ \emph {et~al.}(2010)\citenamefont {Clerk},
  \citenamefont {Devoret}, \citenamefont {Girvin}, \citenamefont {Marquardt},\
  and\ \citenamefont {Schoelkopf}}]{Clerk2010}%
  \BibitemOpen
  \bibfield  {author} {\bibinfo {author} {\bibfnamefont {A.~A.}\ \bibnamefont
  {Clerk}}, \bibinfo {author} {\bibfnamefont {M.~H.}\ \bibnamefont {Devoret}},
  \bibinfo {author} {\bibfnamefont {S.~M.}\ \bibnamefont {Girvin}}, \bibinfo
  {author} {\bibfnamefont {F.}~\bibnamefont {Marquardt}}, \ and\ \bibinfo
  {author} {\bibfnamefont {R.~J.}\ \bibnamefont {Schoelkopf}},\ }\href@noop {}
  {\bibfield  {journal} {\bibinfo  {journal} {Rev. Mod. Phys.}\ }\textbf
  {\bibinfo {volume} {82}},\ \bibinfo {pages} {1155} (\bibinfo {year}
  {2010})}\BibitemShut {NoStop}%
\bibitem [{pur()}]{purity-pres}%
  \BibitemOpen
  \href@noop {} {}\bibinfo {note} {A purity-preserving operation maps a pure
  intial state to a pure final state. A purity-preserving generalized
  measurement maps a pure state to a pure state for any particular measurement
  outcome}\BibitemShut {NoStop}%
\bibitem [{\citenamefont {Korotkov}\ and\ \citenamefont
  {Jordan}(2006)}]{Kor-Jordan-06}%
  \BibitemOpen
  \bibfield  {author} {\bibinfo {author} {\bibfnamefont {A.~N.}\ \bibnamefont
  {Korotkov}}\ and\ \bibinfo {author} {\bibfnamefont {A.~N.}\ \bibnamefont
  {Jordan}},\ }\href@noop {} {\bibfield  {journal} {\bibinfo  {journal} {Phys.
  Rev. Lett.}\ }\textbf {\bibinfo {volume} {97}},\ \bibinfo {pages} {166805}
  (\bibinfo {year} {2006})}\BibitemShut {NoStop}%
\bibitem [{\citenamefont {Jordan}\ and\ \citenamefont
  {Korotkov}(2010)}]{Jordan-Kor-10}%
  \BibitemOpen
  \bibfield  {author} {\bibinfo {author} {\bibfnamefont {A.~N.}\ \bibnamefont
  {Jordan}}\ and\ \bibinfo {author} {\bibfnamefont {A.~N.}\ \bibnamefont
  {Korotkov}},\ }\href@noop {} {\bibfield  {journal} {\bibinfo  {journal}
  {Contemp. Phys.}\ }\textbf {\bibinfo {volume} {51}},\ \bibinfo {pages} {125}
  (\bibinfo {year} {2010})}\BibitemShut {NoStop}%
\bibitem [{\citenamefont {Oreshkov}\ and\ \citenamefont
  {Brun}(2005)}]{Oreshkov2005}%
  \BibitemOpen
  \bibfield  {author} {\bibinfo {author} {\bibfnamefont {O.}~\bibnamefont
  {Oreshkov}}\ and\ \bibinfo {author} {\bibfnamefont {T.~A.}\ \bibnamefont
  {Brun}},\ }\href@noop {} {\bibfield  {journal} {\bibinfo  {journal} {Phys.
  Rev. Lett.}\ }\textbf {\bibinfo {volume} {95}},\ \bibinfo {pages} {110409}
  (\bibinfo {year} {2005})}\BibitemShut {NoStop}%
\bibitem [{\citenamefont {Oreshkov}\ and\ \citenamefont
  {Brun}(2006)}]{Oreshkov2006}%
  \BibitemOpen
  \bibfield  {author} {\bibinfo {author} {\bibfnamefont {O.}~\bibnamefont
  {Oreshkov}}\ and\ \bibinfo {author} {\bibfnamefont {T.~A.}\ \bibnamefont
  {Brun}},\ }\href@noop {} {\bibfield  {journal} {\bibinfo  {journal} {Phys.
  Rev. A}\ }\textbf {\bibinfo {volume} {73}},\ \bibinfo {pages} {042314}
  (\bibinfo {year} {2006})}\BibitemShut {NoStop}%
\bibitem [{\citenamefont {Varbanov}\ and\ \citenamefont
  {Brun}(2007)}]{Varbanov2007}%
  \BibitemOpen
  \bibfield  {author} {\bibinfo {author} {\bibfnamefont {M.}~\bibnamefont
  {Varbanov}}\ and\ \bibinfo {author} {\bibfnamefont {T.~A.}\ \bibnamefont
  {Brun}},\ }\href@noop {} {\bibfield  {journal} {\bibinfo  {journal} {Phys.
  Rev. A}\ }\textbf {\bibinfo {volume} {76}},\ \bibinfo {pages} {032104}
  (\bibinfo {year} {2007})}\BibitemShut {NoStop}%
\bibitem [{\citenamefont {Chow}\ \emph {et~al.}(2013)\citenamefont {Chow},
  \citenamefont {Gambetta}, \citenamefont {Cross}, \citenamefont {Merkel},
  \citenamefont {Rigetti},\ and\ \citenamefont {Steffen}}]{Chow2013}%
  \BibitemOpen
  \bibfield  {author} {\bibinfo {author} {\bibfnamefont {J.~M.}\ \bibnamefont
  {Chow}}, \bibinfo {author} {\bibfnamefont {J.~M.}\ \bibnamefont {Gambetta}},
  \bibinfo {author} {\bibfnamefont {A.~W.}\ \bibnamefont {Cross}}, \bibinfo
  {author} {\bibfnamefont {S.~T.}\ \bibnamefont {Merkel}}, \bibinfo {author}
  {\bibfnamefont {C.}~\bibnamefont {Rigetti}}, \ and\ \bibinfo {author}
  {\bibfnamefont {M.}~\bibnamefont {Steffen}},\ }\href@noop {} {\bibfield
  {journal} {\bibinfo  {journal} {New J. Phys.}\ }\textbf {\bibinfo {volume}
  {15}},\ \bibinfo {pages} {115012} (\bibinfo {year} {2013})}\BibitemShut
  {NoStop}%
\bibitem [{\citenamefont {Dressel}\ and\ \citenamefont
  {Jordan}(2013)}]{Dressel2013b}%
  \BibitemOpen
  \bibfield  {author} {\bibinfo {author} {\bibfnamefont {J.}~\bibnamefont
  {Dressel}}\ and\ \bibinfo {author} {\bibfnamefont {A.~N.}\ \bibnamefont
  {Jordan}},\ }\href@noop {} {\bibfield  {journal} {\bibinfo  {journal} {Phys.
  Rev. A}\ }\textbf {\bibinfo {volume} {88}},\ \bibinfo {pages} {022107}
  (\bibinfo {year} {2013})}\BibitemShut {NoStop}%
\bibitem [{\citenamefont {Gilchrist}\ \emph {et~al.}(2005)\citenamefont
  {Gilchrist}, \citenamefont {Langford},\ and\ \citenamefont
  {Nielsen}}]{Gilchrist2005}%
  \BibitemOpen
  \bibfield  {author} {\bibinfo {author} {\bibfnamefont {A.}~\bibnamefont
  {Gilchrist}}, \bibinfo {author} {\bibfnamefont {N.~K.}\ \bibnamefont
  {Langford}}, \ and\ \bibinfo {author} {\bibfnamefont {M.~A.}\ \bibnamefont
  {Nielsen}},\ }\href@noop {} {\bibfield  {journal} {\bibinfo  {journal} {Phys.
  Rev. A}\ }\textbf {\bibinfo {volume} {71}},\ \bibinfo {pages} {062310}
  (\bibinfo {year} {2005})}\BibitemShut {NoStop}%
\bibitem [{\citenamefont {Nielsen}(2002)}]{Nielsen-02}%
  \BibitemOpen
  \bibfield  {author} {\bibinfo {author} {\bibfnamefont {M.~A.}\ \bibnamefont
  {Nielsen}},\ }\href@noop {} {\bibfield  {journal} {\bibinfo  {journal} {Phys.
  Lett. A}\ }\textbf {\bibinfo {volume} {303}},\ \bibinfo {pages} {249}
  (\bibinfo {year} {2002})}\BibitemShut {NoStop}%
\bibitem [{\citenamefont {Magesan}\ \emph {et~al.}(2011)\citenamefont
  {Magesan}, \citenamefont {Gambetta},\ and\ \citenamefont
  {Emerson}}]{Magesan-11}%
  \BibitemOpen
  \bibfield  {author} {\bibinfo {author} {\bibfnamefont {E.}~\bibnamefont
  {Magesan}}, \bibinfo {author} {\bibfnamefont {J.~M.}\ \bibnamefont
  {Gambetta}}, \ and\ \bibinfo {author} {\bibfnamefont {J.}~\bibnamefont
  {Emerson}},\ }\href@noop {} {\bibfield  {journal} {\bibinfo  {journal} {Phys.
  Rev. Lett.}\ }\textbf {\bibinfo {volume} {106}},\ \bibinfo {pages} {180504}
  (\bibinfo {year} {2011})}\BibitemShut {NoStop}%
\bibitem [{\citenamefont {Kiesel}\ \emph {et~al.}(2005)\citenamefont {Kiesel},
  \citenamefont {Schmid}, \citenamefont {Weber}, \citenamefont {Ursin},\ and\
  \citenamefont {Weinfurter}}]{Kiesel-05}%
  \BibitemOpen
  \bibfield  {author} {\bibinfo {author} {\bibfnamefont {N.}~\bibnamefont
  {Kiesel}}, \bibinfo {author} {\bibfnamefont {C.}~\bibnamefont {Schmid}},
  \bibinfo {author} {\bibfnamefont {U.}~\bibnamefont {Weber}}, \bibinfo
  {author} {\bibfnamefont {R.}~\bibnamefont {Ursin}}, \ and\ \bibinfo {author}
  {\bibfnamefont {H.}~\bibnamefont {Weinfurter}},\ }\href@noop {} {\bibfield
  {journal} {\bibinfo  {journal} {Phys. Rev. Lett.}\ }\textbf {\bibinfo
  {volume} {95}},\ \bibinfo {pages} {210505} (\bibinfo {year}
  {2005})}\BibitemShut {NoStop}%
\bibitem [{\citenamefont {Pedersen}\ \emph {et~al.}(2007)\citenamefont
  {Pedersen}, \citenamefont {Moller},\ and\ \citenamefont
  {Molmer}}]{Pedersen-07}%
  \BibitemOpen
  \bibfield  {author} {\bibinfo {author} {\bibfnamefont {L.~H.}\ \bibnamefont
  {Pedersen}}, \bibinfo {author} {\bibfnamefont {N.~M.}\ \bibnamefont
  {Moller}}, \ and\ \bibinfo {author} {\bibfnamefont {K.}~\bibnamefont
  {Molmer}},\ }\href@noop {} {\bibfield  {journal} {\bibinfo  {journal} {Phys.
  Lett. A}\ }\textbf {\bibinfo {volume} {367}},\ \bibinfo {pages} {47}
  (\bibinfo {year} {2007})}\BibitemShut {NoStop}%
\bibitem [{\citenamefont {Keane}\ and\ \citenamefont
  {Korotkov}(2012)}]{Keane-12}%
  \BibitemOpen
  \bibfield  {author} {\bibinfo {author} {\bibfnamefont {K.}~\bibnamefont
  {Keane}}\ and\ \bibinfo {author} {\bibfnamefont {A.~N.}\ \bibnamefont
  {Korotkov}},\ }\href@noop {} {\bibfield  {journal} {\bibinfo  {journal}
  {Phys. Rev. A}\ }\textbf {\bibinfo {volume} {86}},\ \bibinfo {pages} {012333}
  (\bibinfo {year} {2012})}\BibitemShut {NoStop}%
\bibitem [{\citenamefont {Mi\ifmmode~\check{c}\else \v{c}\fi{}uda}\ \emph
  {et~al.}(2008)\citenamefont {Mi\ifmmode~\check{c}\else \v{c}\fi{}uda},
  \citenamefont {Je\ifmmode~\check{z}\else \v{z}\fi{}ek}, \citenamefont
  {Du\ifmmode~\check{s}\else \v{s}\fi{}ek},\ and\ \citenamefont
  {Fiur\'a\ifmmode~\check{s}\else \v{s}\fi{}ek}}]{Micuda2008}%
  \BibitemOpen
  \bibfield  {author} {\bibinfo {author} {\bibfnamefont {M.}~\bibnamefont
  {Mi\ifmmode~\check{c}\else \v{c}\fi{}uda}}, \bibinfo {author} {\bibfnamefont
  {M.}~\bibnamefont {Je\ifmmode~\check{z}\else \v{z}\fi{}ek}}, \bibinfo
  {author} {\bibfnamefont {M.}~\bibnamefont {Du\ifmmode~\check{s}\else
  \v{s}\fi{}ek}}, \ and\ \bibinfo {author} {\bibfnamefont {J.}~\bibnamefont
  {Fiur\'a\ifmmode~\check{s}\else \v{s}\fi{}ek}},\ }\href@noop {} {\bibfield
  {journal} {\bibinfo  {journal} {Phys. Rev. A}\ }\textbf {\bibinfo {volume}
  {78}},\ \bibinfo {pages} {062311} (\bibinfo {year} {2008})}\BibitemShut
  {NoStop}%
\bibitem [{\citenamefont {Bongioanni}\ \emph {et~al.}(2010)\citenamefont
  {Bongioanni}, \citenamefont {Sansoni}, \citenamefont {Sciarrino},
  \citenamefont {Vallone},\ and\ \citenamefont {Mataloni}}]{Bongioanni2010}%
  \BibitemOpen
  \bibfield  {author} {\bibinfo {author} {\bibfnamefont {I.}~\bibnamefont
  {Bongioanni}}, \bibinfo {author} {\bibfnamefont {L.}~\bibnamefont {Sansoni}},
  \bibinfo {author} {\bibfnamefont {F.}~\bibnamefont {Sciarrino}}, \bibinfo
  {author} {\bibfnamefont {G.}~\bibnamefont {Vallone}}, \ and\ \bibinfo
  {author} {\bibfnamefont {P.}~\bibnamefont {Mataloni}},\ }\href@noop {}
  {\bibfield  {journal} {\bibinfo  {journal} {Phys. Rev. A}\ }\textbf {\bibinfo
  {volume} {82}},\ \bibinfo {pages} {042307} (\bibinfo {year}
  {2010})}\BibitemShut {NoStop}%
\bibitem [{\citenamefont {Wootters}(1981)}]{Wootters1981}%
  \BibitemOpen
  \bibfield  {author} {\bibinfo {author} {\bibfnamefont {W.~K.}\ \bibnamefont
  {Wootters}},\ }\href@noop {} {\bibfield  {journal} {\bibinfo  {journal}
  {Phys. Rev. D}\ }\textbf {\bibinfo {volume} {23}},\ \bibinfo {pages} {357}
  (\bibinfo {year} {1981})}\BibitemShut {NoStop}%
\bibitem [{\citenamefont {Uhlmann}(1976)}]{Uhlmann1976}%
  \BibitemOpen
  \bibfield  {author} {\bibinfo {author} {\bibfnamefont {A.}~\bibnamefont
  {Uhlmann}},\ }\href@noop {} {\bibfield  {journal} {\bibinfo  {journal} {Rep.
  Math. Phys.}\ }\textbf {\bibinfo {volume} {9}},\ \bibinfo {pages} {273}
  (\bibinfo {year} {1976})}\BibitemShut {NoStop}%
\bibitem [{\citenamefont {Jozsa}(1994)}]{Jozsa-94}%
  \BibitemOpen
  \bibfield  {author} {\bibinfo {author} {\bibfnamefont {R.}~\bibnamefont
  {Jozsa}},\ }\href@noop {} {\bibfield  {journal} {\bibinfo  {journal} {J. Mod.
  Opt.}\ }\textbf {\bibinfo {volume} {41}},\ \bibinfo {pages} {2315} (\bibinfo
  {year} {1994})}\BibitemShut {NoStop}%
\bibitem [{\citenamefont {Barnum}\ \emph {et~al.}(1996)\citenamefont {Barnum},
  \citenamefont {Caves}, \citenamefont {Fuchs}, \citenamefont {Jozsa},\ and\
  \citenamefont {Schumacher}}]{Barnum-96}%
  \BibitemOpen
  \bibfield  {author} {\bibinfo {author} {\bibfnamefont {H.}~\bibnamefont
  {Barnum}}, \bibinfo {author} {\bibfnamefont {C.~M.}\ \bibnamefont {Caves}},
  \bibinfo {author} {\bibfnamefont {C.~A.}\ \bibnamefont {Fuchs}}, \bibinfo
  {author} {\bibfnamefont {R.}~\bibnamefont {Jozsa}}, \ and\ \bibinfo {author}
  {\bibfnamefont {B.}~\bibnamefont {Schumacher}},\ }\href@noop {} {\bibfield
  {journal} {\bibinfo  {journal} {Phys. Rev. Lett.}\ }\textbf {\bibinfo
  {volume} {76}},\ \bibinfo {pages} {2818} (\bibinfo {year}
  {1996})}\BibitemShut {NoStop}%
\bibitem [{\citenamefont {Jamio{\l}kowski}(1974)}]{Jamiolkowski1974}%
  \BibitemOpen
  \bibfield  {author} {\bibinfo {author} {\bibfnamefont {A.}~\bibnamefont
  {Jamio{\l}kowski}},\ }\href@noop {} {\bibfield  {journal} {\bibinfo
  {journal} {Rep. Math. Phys.}\ }\textbf {\bibinfo {volume} {5}},\ \bibinfo
  {pages} {415} (\bibinfo {year} {1974})}\BibitemShut {NoStop}%
\bibitem [{\citenamefont {Choi}(1975)}]{Choi1975}%
  \BibitemOpen
  \bibfield  {author} {\bibinfo {author} {\bibfnamefont {M.-D.}\ \bibnamefont
  {Choi}},\ }\href@noop {} {\bibfield  {journal} {\bibinfo  {journal} {Linear
  Alg. Appl.}\ }\textbf {\bibinfo {volume} {10}},\ \bibinfo {pages} {285}
  (\bibinfo {year} {1975})}\BibitemShut {NoStop}%
\bibitem [{\citenamefont {Jiang}\ \emph {et~al.}(2013)\citenamefont {Jiang},
  \citenamefont {Luo},\ and\ \citenamefont {Fu}}]{Min2013}%
  \BibitemOpen
  \bibfield  {author} {\bibinfo {author} {\bibfnamefont {M.}~\bibnamefont
  {Jiang}}, \bibinfo {author} {\bibfnamefont {S.}~\bibnamefont {Luo}}, \ and\
  \bibinfo {author} {\bibfnamefont {S.}~\bibnamefont {Fu}},\ }\href@noop {}
  {\bibfield  {journal} {\bibinfo  {journal} {Phys. Rev. A}\ }\textbf {\bibinfo
  {volume} {87}},\ \bibinfo {pages} {022310} (\bibinfo {year}
  {2013})}\BibitemShut {NoStop}%
\bibitem [{\citenamefont {Raginsky}(2001)}]{Raginsky2001}%
  \BibitemOpen
  \bibfield  {author} {\bibinfo {author} {\bibfnamefont {M.}~\bibnamefont
  {Raginsky}},\ }\href@noop {} {\bibfield  {journal} {\bibinfo  {journal}
  {Phys. Lett. A}\ }\textbf {\bibinfo {volume} {290}},\ \bibinfo {pages} {11}
  (\bibinfo {year} {2001})}\BibitemShut {NoStop}%
\bibitem [{\citenamefont {Korotkov}\ and\ \citenamefont
  {Keane}(2010)}]{Kor-Keane-10}%
  \BibitemOpen
  \bibfield  {author} {\bibinfo {author} {\bibfnamefont {A.~N.}\ \bibnamefont
  {Korotkov}}\ and\ \bibinfo {author} {\bibfnamefont {K.}~\bibnamefont
  {Keane}},\ }\href@noop {} {\bibfield  {journal} {\bibinfo  {journal} {Phys.
  Rev. A}\ }\textbf {\bibinfo {volume} {81}},\ \bibinfo {pages} {040103}
  (\bibinfo {year} {2010})}\BibitemShut {NoStop}%
\end{thebibliography}

\end{document}